\documentclass{PoS}
\usepackage{texnames}
\usepackage[T1]{fontenc}
\usepackage{cite}
\usepackage{lineno}
\usepackage{url}
\usepackage{subcaption}

\title{Future heavy-ion facilities: FCC-AA}

\ShortTitle{Future heavy-ion facilities: FCC-AA}

\author{\speaker{A.~Dainese}\,$^a$, L.~Apolin\'ario\,$^b$,  N.~Armesto\,$^c$, D.~d'Enterria\,$^d$, J.M.~Jowett\,$^d$, J.-P.~Lansberg\,$^e$, J.G.~Milhano\,$^{b,d}$, C.A.~Salgado\,$^c$, M.~Schaumann\,$^d$, M.~van~Leuween\,$^{f,g,d}$, U.A.~Wiedemann\,$^d$\\~\\
        $^a$~INFN - Sezione di Padova, Padova, Italy\\$^b$~LIP, Lisboa, Portugal\\ $^c$~Instituto Galego de F\'{\i}sica de Altas Enerx\'{\i}as, Universidade de Santiago de Compostela, Galicia-Spain\\$^d$~CERN, Geneva, Switzerland\\$^e$~IPNO, Univ. Paris-Sud, CNRS/IN2P3, Universit\'e Paris-Saclay, Orsay, France\\ $^f$~Utrecht University, the Netherlands\\$^g$~NIKHEF, the Netherlands.\\
        E-mail: \email{andrea.dainese@pd.infn.it}}

\abstract{The operation of the Future Circular Collider (FCC) with heavy ions 
would provide Pb--Pb and p--Pb collisions at
$\sqrt{s_{\mathrm{\scriptscriptstyle NN}}}=39$ and 63~TeV, respectively, per nucleon--nucleon
collision, with projected per-month integrated luminosities of up to
110~nb$^{-1}$ and 29~pb$^{-1}$, respectively.
This document outlines the unique and broad physics opportunities with heavy ions at the energy frontier opened by FCC.}

\FullConference{International Conference on Hard and Electromagnetic Probes of High-Energy Nuclear Collisions\\
		30 September - 5 October 2018\\
		Aix-Les-Bains, Savoie, France}

\def \gsim{\mathrel{\vcenter
     {\hbox{$>$}\nointerlineskip\hbox{$\sim$}}}}
\def \lsim{\mathrel{\vcenter
     {\hbox{$<$}\nointerlineskip\hbox{$\sim$}}}}

\newcommand{\PbPb}{{Pb--Pb}}

\newcommand{\sqrtsNN}{\ensuremath{\sqrt{s_{\mathrm{\scriptscriptstyle NN}}}}}

\newcommand{\jpsi}{\ensuremath{J/\psi}}

\newcommand{\bbbar}{\ensuremath{{b\overline{b}}}}

\newcommand{\RAA}{\ensuremath{R_{\mathrm{AA}}}}

\newcommand{\gev}{\ensuremath{\mathrm{GeV}}}
\newcommand{\tev}{\ensuremath{\mathrm{TeV}}}

\newcommand{\GeVc}{\ensuremath{\mathrm{~GeV}/c}}

\newcommand{\beq}{\begin{equation}}
\newcommand{\eeq}{\end{equation}}
\newcommand{\beqn}{\begin{eqnarray}}
\newcommand{\eeqn}{\end{eqnarray}}

\newcommand{\beqa}{\begin{eqnarray}}
\newcommand{\eeqa}{\end{eqnarray}}
\def\lsim{\raise0.3ex\hbox{$<$\kern-0.75em\raise-1.1ex\hbox{$\sim$}}}
\def\gsim{\raise0.3ex\hbox{$>$\kern-0.75em\raise-1.1ex\hbox{$\sim$}}}

\newcommand{\pt}{\ensuremath{p_{\mathrm{T}}}}

\newcommand{\sqrtsnn}{\sqrtsNN}

\newcommand{\ttbar}    {t\bar{t}}

\providecommand{\PbPb}{Pb--Pb}

\providecommand{\gp}{\gamma\,p}

\newcommand*{\cm}{c.m.\@}

\begin{document}

\section{Introduction}
\label{sec:HI_intro}

The operation of the FCC-hh (hadron--hadron) with heavy-ion beams is part of the accelerator design studies~\cite{Benedikt:2651300,Mangano:2651294}.
For a centre-of-mass energy $\sqrt{s}= 100$~TeV for pp collisions, the relation $\sqrt{s_{\rm NN}}= \sqrt {s} \sqrt{Z_1 Z_2 / A_1 A_2}$ 
gives the energy per nucleon--nucleon collision of $\sqrt{s_{\rm NN}} = 39.4$~TeV for Pb--Pb ($Z=82$, $A=208$) and 62.8~TeV for p--Pb collisions. 
The present estimate of the integrated luminosity per month of running is larger by factors 10--30 than the current projection for the future LHC runs~\cite{Citron:2018lsq}. 
The possibility of using nuclei smaller than Pb, like e.g. $^{40}$Ar, $^{84}$Kr or $^{129}$Xe, to achieve larger instantaneous luminosity is also under consideration. 

The significant increase in the centre-of-mass energy and integrated luminosity with respect to the LHC opens up novel opportunities for physics studies of the Quark-Gluon Plasma (QGP) and of gluon saturation, among others, described in a recent CERN Yellow Report~\cite{Dainese:2016gch} and in the volume on Physics Opportunities of the FCC Conceptual Design Report~\cite{Mangano:2651294}. 
The main scientific motivations to carry out measurements with heavy ions at the FCC are discussed in this document.

\section{Summary of the heavy-ion performance of FCC-hh}
\label{sec:HI_machine}

It has been shown that the FCC-hh could operate very efficiently as a nucleus-nucleus or proton-nucleus collider, analogously to the LHC. Previous studies~\cite{Schaumann:2015fsa,Dainese:2016gch} have revealed that it enters a new, highly-efficient operating regime, in which a large fraction of the injected intensity can be converted to useful integrated luminosity. 
Table~\ref{tab:IonParametersShortForPhysics} summarises the key parameters
for Pb--Pb and p--Pb operation. 
Two beam parameter cases have been considered,
\textit{baseline} and \textit{ultimate}, which differ in the $\beta$-function at the interaction point, the optical function $\beta^*$ at the interaction point, and the assumed bunch spacing, defining the maximum number of circulating bunches.
The luminosity is shown for one experiment but the case of two experiments was also studied:
this decreases the integrated luminosity per experiment by 40\%, but increases the total by 20\%.   
The performance projections assume the LHC to be the final injector synchrotron before the FCC~\cite{Citron:2018lsq}. 
A performance efficiency factor was taken into account to include set-up time, early beam aborts and other deviations from the idealised running on top of the theoretical calculations. 
Further details on the performance of the heavy-ion operation in FCC-hh can be found in Ref.~\cite{Benedikt:2651300}.

\begin{table}[!h]
\begin{center}
\caption[Main beam and collider parameters.]{Beam and machine parameters.} 
\label{tab:IonParametersShortForPhysics}%
\begin{tabular}{lc|cc|cc}
\hline   
			&	Unit	& \multicolumn{2}{c|}{Baseline} & \multicolumn{2}{c}{Ultimate} \\
\hline
Operation mode&-						& Pb--Pb 	& p--Pb 	& Pb--Pb & p--Pb \\

Number of Pb bunches & -    					& \multicolumn{2}{c|}{2760} 	& \multicolumn{2}{c}{5400} \\

Bunch spacing & [ns]					& \multicolumn{2}{c|}{100} & \multicolumn{2}{c}{50}\\

Peak luminosity (1 experiment) & [$10^{27}\mathrm{cm}^{-2}\mathrm{s}^{-1}$]    & 80 		& 13300 & 320 & 55500 \\

Integrated luminosity (1 experiment, 30 days) & [nb$^{-1}$] & 35	&   8000	& 110 & 29000\\

\hline

\end{tabular}%
\end{center}
\end{table}%

\section{Global characteristics of Pb--Pb collisions}
\label{sec:HI_global}

Extrapolating measurements of charged particle multiplicity, transverse energy and femtoscopic correlations 
at lower energies, one can obtain estimates
for the growth of global event characteristics from LHC to FCC. In particular, up to the top LHC energy, the growth of charged hadron 
event multiplicity per unit rapidity in Pb--Pb collisions is consistent with a slowly-rising power-law:
   ${\rm d}N_{\rm ch}/{\rm d}\eta\,(\eta=0) \propto (\sqrtsNN)^{0.3}$.
As shown in Table~\ref{tab:PbPb}, this amounts to an increase of a factor $\sim 1.8$ from LHC to FCC. 

\begin{table}[h]
\caption{Global properties measured in central Pb--Pb collisions (0--5\% centrality class) at $\sqrtsNN=2.76$~TeV and extrapolated to 5.5 and 39~TeV.
}
\small
\begin{center}
\begin{tabular}{lccc}
\hline
Quantity & Pb--Pb 2.76~TeV & Pb--Pb 5.5~TeV & Pb--Pb 39~TeV \\
\hline
${\rm d}N_{\rm ch}/{\rm d}\eta$ at $\eta=0$ & 1600 & 2000 & 3600 \\
${\rm d}E_{\rm T}/{\rm d}\eta$ at $\eta=0$ & 1.8--2.0~TeV & 2.3--2.6~TeV & 5.2--5.8~TeV \\
Homogeneity volume $$ & 5000~fm$^3$  & 6200~fm$^3$ & 11000~fm$^3$ \\
Decoupling time & 10~fm/$c$ &  11~fm/$c$ & 13~fm/$c$ \\
$\varepsilon$ at $\tau=1$~fm/$c$ & 12--13~GeV/fm$^3$  & 16--17~GeV/fm$^3$ & 35--40~GeV/fm$^3$ \\
\hline
\end{tabular}
\end{center}
\label{tab:PbPb}
\end{table}

\begin{figure}[!t]
\begin{center}
\includegraphics[width=0.45\textwidth]{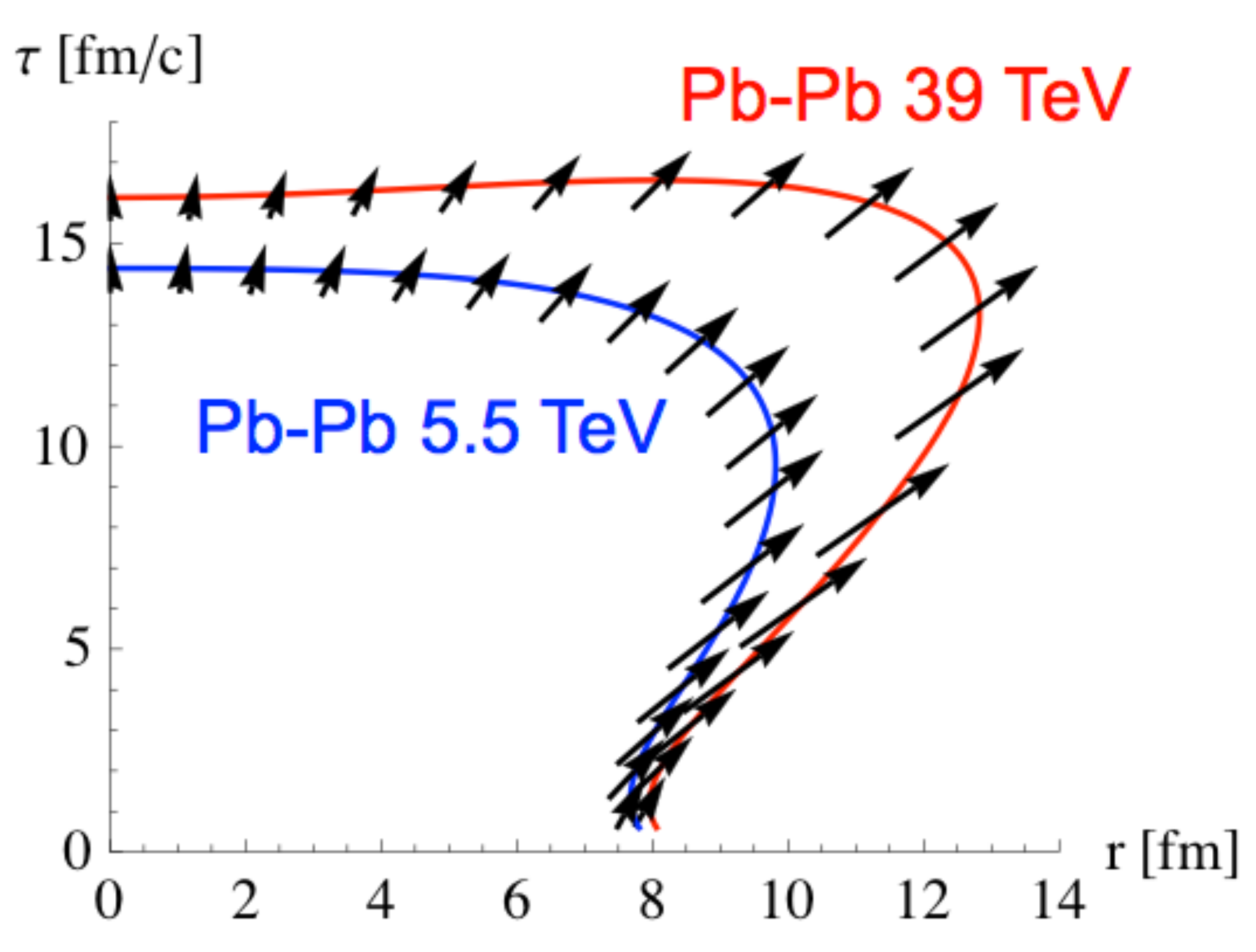}
\includegraphics[width=0.42\textwidth]{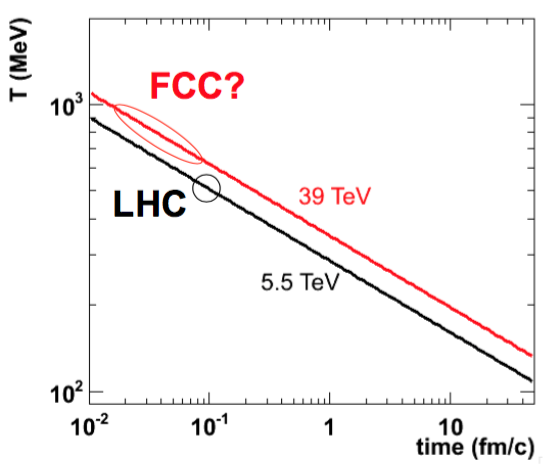}
\caption{Left: Space-time profile at freeze-out from hydrodynamical calculations for central Pb--Pb collisions at $\sqrtsNN=5.5$~TeV and $39$~TeV.
Right: Time evolution of the QGP temperature as estimated on the basis of the Bjorken relation and the Stefan-Boltzmann equation (see text for details).}
\label{fig:freezeout}
\end{center}
\end{figure}

Figure~\ref{fig:freezeout} (left) shows results for the freeze-out hypersurfaces of central Pb--Pb collisions at different collision energies. This figure quantifies the expectation 
that the denser system created at higher collision energy has to expand to a larger volume and for a longer time before reaching the freeze-out temperature at which decoupling from a partonic to hadronic phase takes place.
The arrows overlaid with the freeze-out hypersurface in Fig.~\ref{fig:freezeout} (left) indicate the transverse flow of the fluid element at decoupling. This provides quantitative support for 
the qualitative expectation that in a larger and more long-lived system, collective effects can grow stronger. 
In general, the global event characteristics listed in Table~\ref{tab:PbPb} determine the spatio-temporal extent of the ``cauldron'' in which QCD matter is evolved, and they
constrain the thermodynamic conditions that apply after thermalization. The measured transverse energy per unit rapidity ${\rm d}E_{\rm T}/{\rm d}\eta$ (see Table~\ref{tab:PbPb})
is of particular importance since it constrains the initial energy density. 
The energy density is expected to increase by a factor of two from LHC to FCC, reaching a value of 35--40~GeV/fm$^3$ at the time of 1~fm/$c$~\cite{Dainese:2016gch}. 
 In Fig.~\ref{fig:freezeout} (right), we have plotted the time-dependence of the QGP plasma temperature for Pb--Pb collisions at the LHC and at the FCC. 
The figure shows that while the temperature increase 
at a given time is a modest 30\% when going from LHC to FCC, the thermalization time of the system is expected to be significantly smaller. 
One may reach initial temperatures as large as $T_0\approx800$--$1000$~MeV at the time $\mathcal{O}(0.02\, {\rm fm}/c)$ after which both nuclei traverse each other at FCC energies.

\section{QGP studies: hard probes}

\label{sec:HI_hardprobes}

\subsection{Jet quenching}
\label{sec:HI_quenching}

The modification of jet properties in heavy-ion collisions with respect to the proton--proton case, commonly referred to as jet quenching, results from the interaction of jet constituents with the QGP that they traverse. 
Over the last few years, as several jet properties were measured in heavy-ion collisions,
the theoretical understanding of jet--QGP interactions has evolved from the early descriptions of single parton energy loss
towards an overall understanding of how full jets are modified by the QGP.
The increase in centre-of-mass energy, the abundance of probes, especially those involving electroweak bosons together with jets, and the qualitatively new processes available (e.g.\,boosted systems, see below) make of the FCC-hh the best-suited future machine for a deeper understanding of this physics.

\subsubsection{Hard processes at FCC-hh energies}
\label{sec:HI_xsections}

\begin{figure}[!t]
\begin{center}
\includegraphics[width=0.5\textwidth]{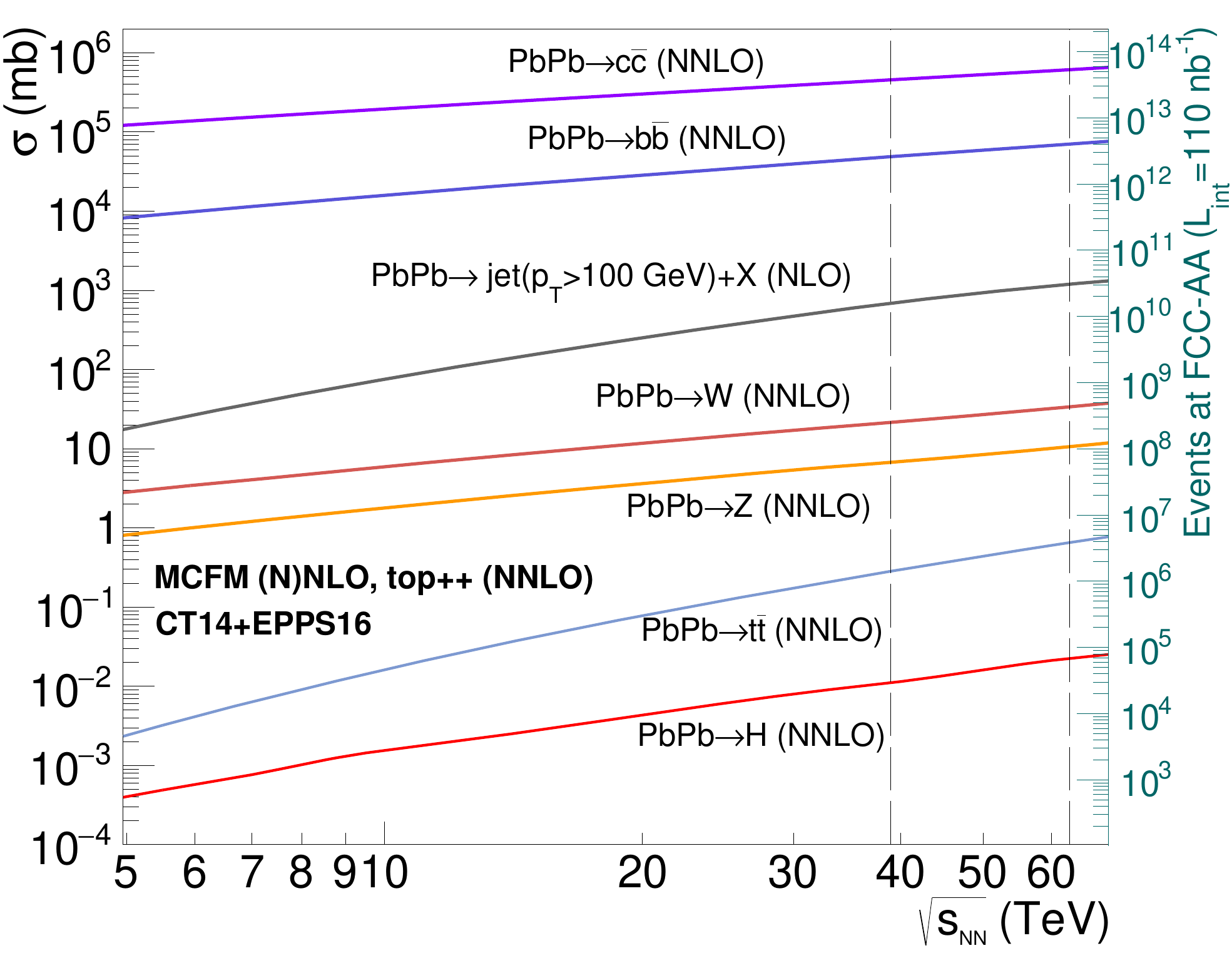}
\hfill
\includegraphics[width=0.4\columnwidth]{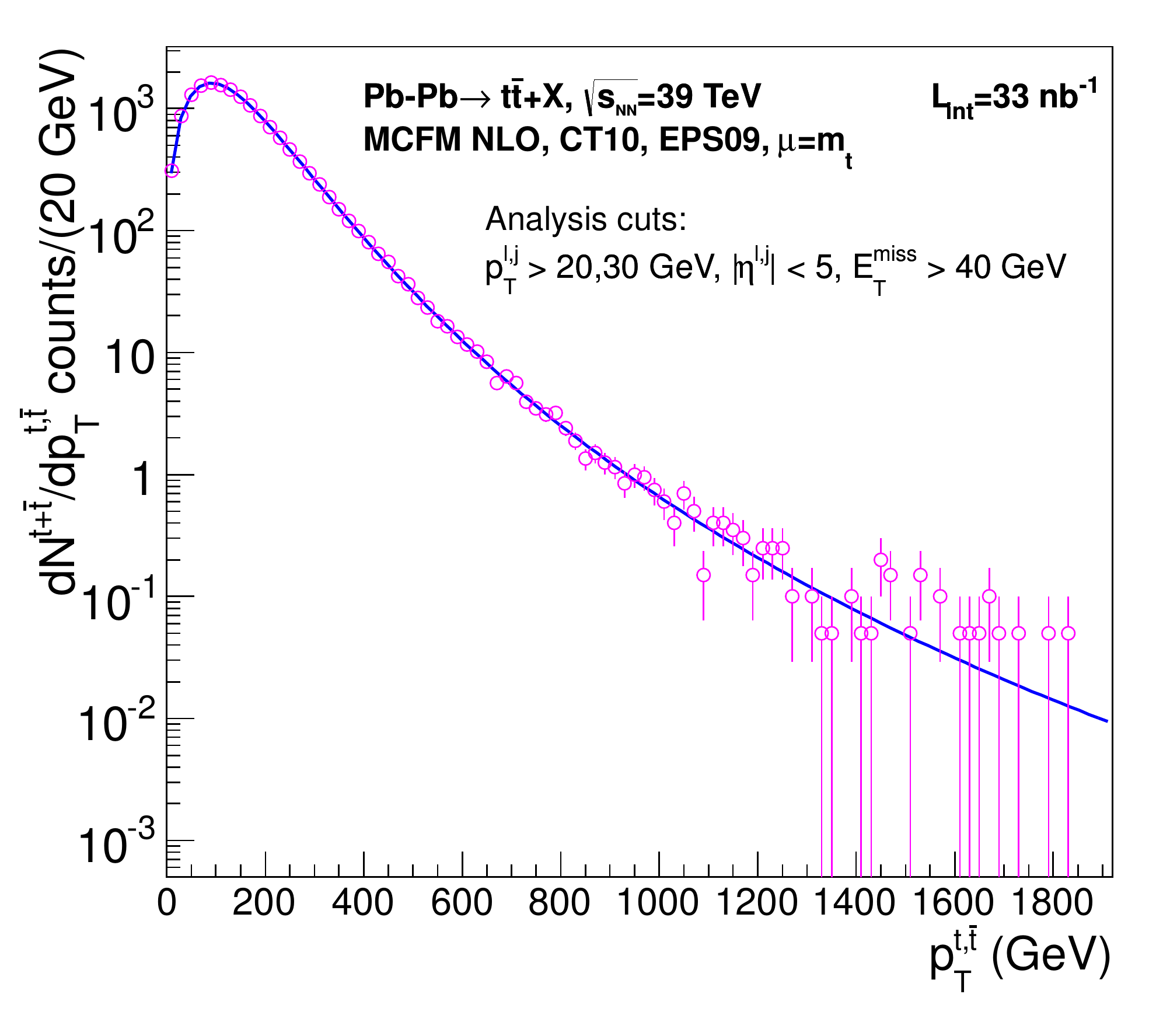}
\caption{Left: Centre-of-mass energy dependence of the cross sections for hard processes of
  interest for a heavy-ion programme, calculated with MCFM~\cite{Campbell:2010ff}
at NNLO and top++~\cite{Czakon:2013goa}
(for all heavy-quarks~\cite{dEnterria:2016ids}).
  Right: Expected top-quark $\pt$ distribution in \PbPb\ in the
  fully-leptonic decay modes at $\sqrtsnn=39$~TeV after acceptance and
  efficiency cuts with the statistical uncertainties
  for the baseline scenario $L_{\rm int}=33$~nb$^{-1}$ (adapted from~\cite{dEnterria:2015mgr}).}
\label{fig:hardXsectHIC}
\end{center}
\end{figure}

The large increase in energy and luminosity from the LHC to the FCC provides new tools to study the matter created in the collisions of heavy ions.
In Fig.~\ref{fig:hardXsectHIC} (left), cross sections for different processes
and different energies are computed with  MCFM~\cite{Campbell:2010ff}
at next-to-next-to-leading order (NNLO) and top++~\cite{Czakon:2013goa}
(for all heavy-quarks~\cite{dEnterria:2016ids}). The increases amount to a factor $\sim
6$ for beauty production, $\sim 10$ for $Z$ production, $\sim 20$ for $Z$+jet and $\sim 80$
for top production. 
The large yields in $Z+$jets (several tens of millions) will enable the study the jet quenching process with excellent calibration of the jet energy. In principle, the measurement of  the energy lost by the jet in $Z$+jet provide a good experimental control of the distribution of the parton energy losses in hot QCD matter.

The motivations for measurements of top quarks in heavy-ion collisions at FCC are multifold. For example, in
p--Pb collisions the cross sections efficiently probe the nuclear
gluon PDFs in a wide range in momentum fraction
$x$ at high scale $Q\sim m_{\rm t}$~\cite{dEnterria:2015mgr} (see Section~\ref{sec:HI_smallxhadro}). In Pb--Pb
collisions, the top-quark observables are sensitive to the energy-loss of heavy quarks~\cite{Baskakov:2015nxa}
and by selecting boosted (very high-$p_{\rm T}$) top quarks one can probe the space-time evolution of jet quenching in the QGP as the decays of boosted top quarks get Lorentz
time dilated (see next section). 
For example, the estimated measurable yields (using
 per-month luminosities from Section~\ref{sec:HI_machine}) with
realistic analysis cuts 
and conservative 50\% efficiency for $b$-jet tagging are 3--$10\times 10^5$ pairs (baseline--ultimate) 
in Pb--Pb collisions for $\ttbar\to \bbbar\,\ell\ell\,\nu\nu$.

As mentioned above, the $p_{\rm T}$ reach of top quarks in Pb--Pb collisions is of special importance for QGP
studies. Figure~\ref{fig:hardXsectHIC} (right) shows the estimated $p_{\rm T}$ spectrum
of the yields (per year) in Pb--Pb collisions for top-quark pair production. The figure indicates that one
could measure top quarks approximately up to $p_{\rm T} \approx
1.8~{\rm TeV}/c$, even considering only one run in the baseline
luminosity scenario. At mid-rapidity, $p_{\rm T}$ as large
as this would correspond approximately to a factor of 10 time dilation
in the top decay (see next section).

Another potential novel probe of the QGP medium at FCC energies is the
Higgs boson. The Higgs boson has a lifetime of $\tau\approx 50$~fm/$c$, which is much larger than the 
time extent of the QGP phase~\cite{dEnterria:2018bqi,Berger:2018mtg}. 
In Ref.~\cite{dEnterria:2018bqi}
it has been argued that the Higgs boson
interacts strongly with the quarks and gluons of the QGP and the interactions
induce its decay in the gluon--gluon or quark--antiquark channels, thus depleting the 
branching ratio to the most common ``observation'' channels $\gamma\gamma$ or $\rm ZZ^\star$.
More recent detailed theoretical calculations, including virtual corrections, predict however no visible 
suppression of the scalar boson~\cite{Ghiglieri:2019lzz}.
The cross section for Higgs boson production in Pb--Pb collisions is expected to
increase by a factor larger than 20 when going from
$\sqrtsNN=5.5~\tev$ to $\sqrtsNN=39~\tev$~\cite{dEnterria:2017jyt}.
The statistical significance of the Higgs boson observation in the
$\gamma\gamma$ decay channel in one Pb--Pb run
at FCC-hh was estimated to be 5.5 (9.5) $\sigma$ in the baseline
(ultimate) luminosity scenarios~\cite{dEnterria:2017jyt}.
The analysis used similar photon selections as used by ATLAS and CMS in pp collisions: 
$\pt > 30,40$~GeV/$c$, $|\eta| < 4$, $R_{\rm isol} = 0.3$. The backgrounds included the irreducible QCD diphoton continuum plus 30\% of events coming from misidentified 
$\gamma$-jet and jet-jet processes. 

\subsubsection{Boosted tops and the space-time picture of the QGP}
\label{sec:HI_boostedtops}

The  FCC will provide large rates of highly-boosted heavy particles, such as tops, $Z$ and $W$ bosons. It is expected that when these particles decay the density profile of the QGP has already evolved. It has been argued that the hadronically-decaying $W$ bosons in events with a $t\bar t$ pair can provide unique insights into the time structure of the QGP~\cite{Apolinario:2017sob}. This is because the time decays of the top and the $W$ bosons are followed by a time-delay in the interaction of the decay products of the $W$ boson with the surrounding medium due to a color coherence effect. The sum of the three times, that reaches values of several fm/$c$ for boosted tops, would be the time at which the interaction with the QGP begins, providing a unique way to directly measure the time structure of the QGP evolution.
In addition, energy loss  would be initially absent for the colour-singlet $q\overline q$ decay products of a highly-boosted $W$ boson: the two quarks would start to be quenched only when their distance becomes larger than the colour correlation length of the medium, which depends on the transport coefficient $\hat{q}$ (the average transverse momentum squared that particles exchange with the medium per unit mean-free path)~\cite{CasalderreySolana:2012ef}.

\begin{figure}[!t]
\begin{center}
\includegraphics[width=0.5\textwidth]{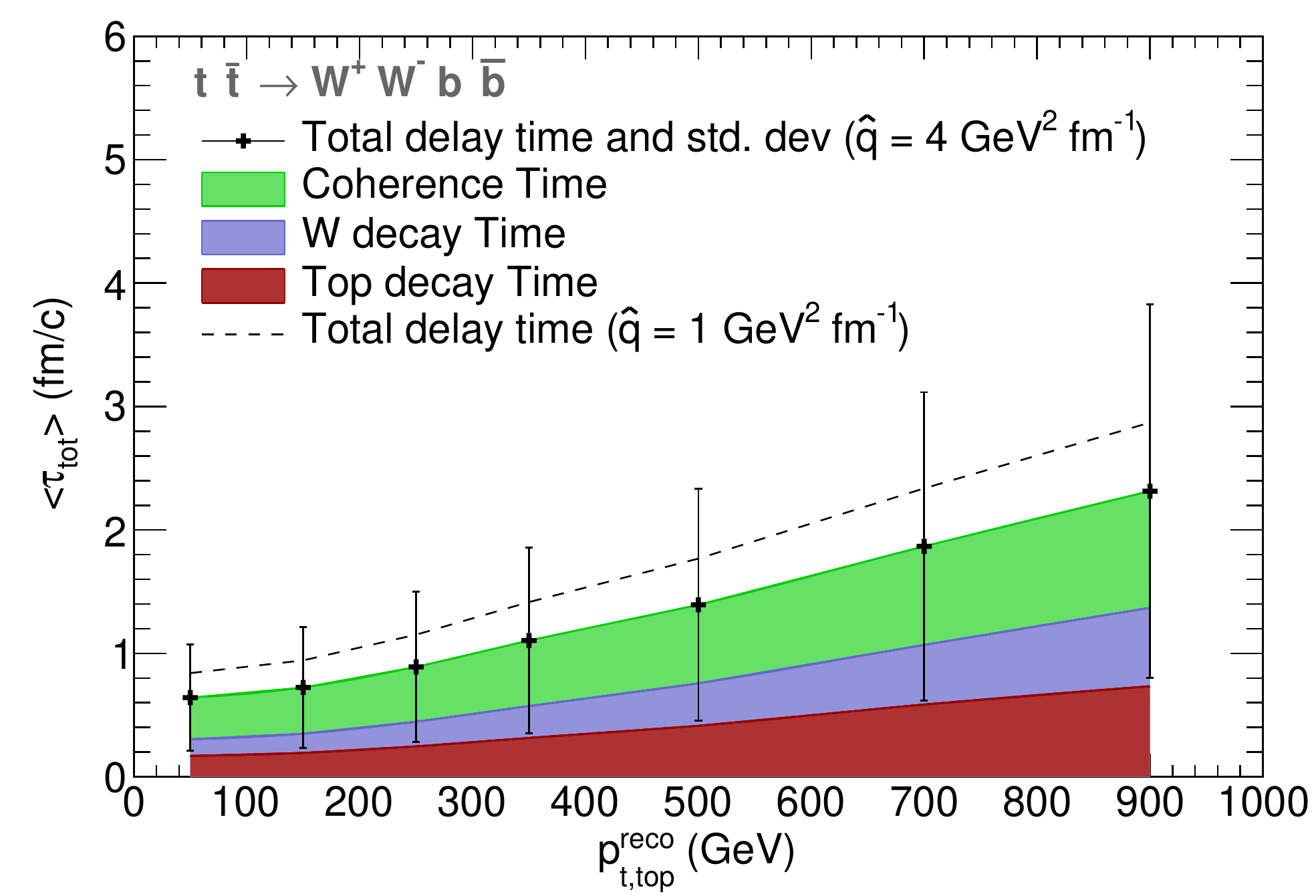}
\includegraphics[width=0.46\textwidth]{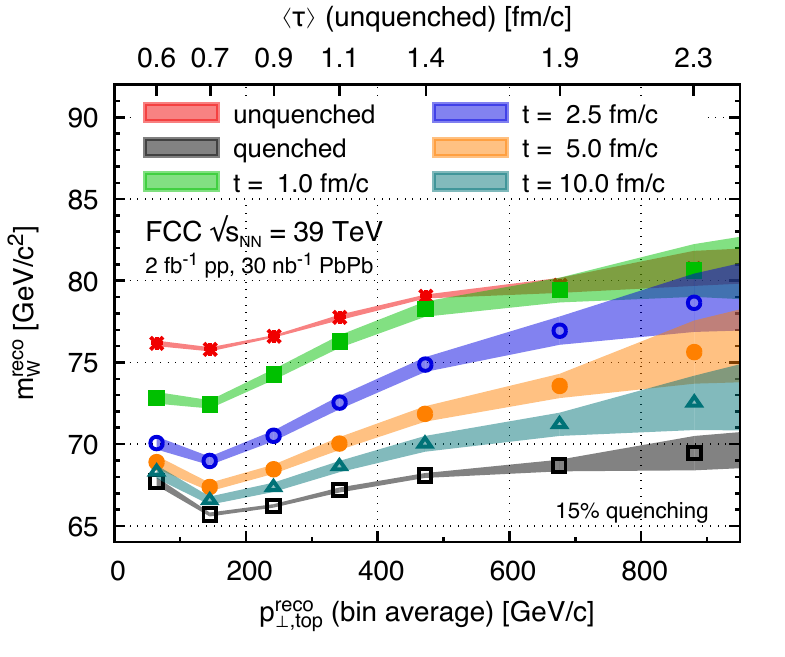}
\caption{Left: Total delay time for $\hat{q} = 4$~GeV$^2$/fm as a function of the top transverse momentum (black dots) and its standard deviation (error bars). The average contribution of each component is shown as a coloured stack band. The dashed line corresponds to a $\hat{q} = 1$~GeV$^2$/fm. Right: Reconstructed $W$ boson mass at FCC energies $\sqrtsNN = 30$~TeV, as a function of the top $\pt$. The upper axis refers to the average total time delay of the corresponding top $\pt$ bin.  Figures from~\cite{Apolinario:2017sob}.}
\label{fig:tops}
\end{center}
\end{figure}

The effect on the reconstructed masses of the top and $W$ is studied, for $t\bar t$ events decaying semileptonically, with different energy loss scenarios as a proof of concept of the potential of these observables to access completely novel quantities in heavy-ion collisions. As shown in Fig.~\ref{fig:tops} (left), times in the range $0.3$--$3$~fm/$c$ are obtained when adding the time delay from Lorentz boosts of the decaying top quark and $W$ boson and the time in which a singlet antenna remains in a colour coherent state.

The reconstructed $W$ jet mass as a function of top transverse momentum at $\sqrtsNN = 39$~TeV is shown in Fig.~\ref{fig:tops} (right). For details on the simulation and reconstruction procedure see~\cite{Apolinario:2017sob}. The shaded region corresponds to the statistical uncertainty obtained with a bootstrap analysis considering central Pb--Pb collisions for $L_{\rm int} = 30$~nb$^{-1}$
 for the Pb--Pb energy loss scenarios and $L_{\rm int} = 2$~fb$^{-1}$ for the pp reference.
Energy loss was simulated considering that all particles, except the $W$ boson decay products, lose $15\%$ of their initial momentum. 
Average time delays $\tau_m = 1$, 2.5, 5, and 10~fm$/c$ were considered as effective QGP time evolution profiles. 
Fig.~\ref{fig:tops} (right) shows a clear separation between the totally quenched case and those in which the time-delay effectively reduces the interaction with the QGP and hence the quenching, especially for the highest boosts considered, that will be accessible at the FCC. This large sensitivity will only be possible at the FCC and partially at the HE-LHC, while only marginally with the present LHC programme~\cite{Apolinario:2017sob}.

\subsection{Heavy flavour and quarkonia}
\label{sec:HI_hf}

Heavy quarks (charm and bottom) are among the hard probes that have
provided important insights on the formation and the characterics of
the QGP~\cite{Andronic:2015wma}.
On the one hand, quarkonium states are sensitive to the formation and to the
temperature of a deconfined plasma via the mechanism of colour-charge
screening, which is thought to be to some extent balanced by the
recombination of heavy quarks and antiquarks from the plasma.
On the other hand, the production of hadrons with open heavy flavour
is sensitive to the QGP-induced modification of the momentum value and
direction of heavy quarks, thus providing
insight on the interaction mechanisms of heavy quarks
with the constituents of the QGP and on its
transport properties.

Here we focus on a few selected aspects that could
represent novel or particularly remarkable observations at FCC energy,
namely: i) large production of thermal charm from
  interactions of light quarks and gluons within the QGP;
ii) observation of an enhancement of charmonium production with
  respect to the binary scaling of the yields in pp collisions, as
  consequence of (re)generation;
iii) observation of a colour screening and (re)generation for the most tightly-bound
  quarkonium state, the $\Upsilon$(1S).

\subsubsection{Thermal charm production}

Interactions between gluons or light quarks of the QGP can lead to the
production of $c\overline c$ pairs if the energy in the centre of mass 
of the interaction is of the order of twice the charm quark mass
$\sqrt{\hat s}\sim 2\,m_c\sim 3$~GeV. 
In Section~\ref{sec:HI_global} we have estimated 
that an initial temperature $T_0$ larger than 800~MeV could be
reached at FCC. 
In Ref.~\cite{Zhou:2016wbo} a detailed hydrodynamical calculation gives 
$T_0=580~$MeV at initial time $\tau_0=0.6$~fm/$c$ for LHC ($\sqrtsNN=5.5~\tev$) and
$T_0=840~$MeV at $\tau_0=0.3$~fm/$c$ for FCC.
With these average QGP temperatures, a sizeable fraction of the (Bose-Einstein distributed) gluons and (Fermi-Dirac distributed) light quarks
have energies larger than the charm quark mass 
and $c\overline c$ pairs can be produced in their interactions. This
production is concentrated in the initial  $\sim 1$~fm/$c$ of the QGP evolution.

\begin{figure}[!t]
\begin{center}
\includegraphics[width=0.42\textwidth]{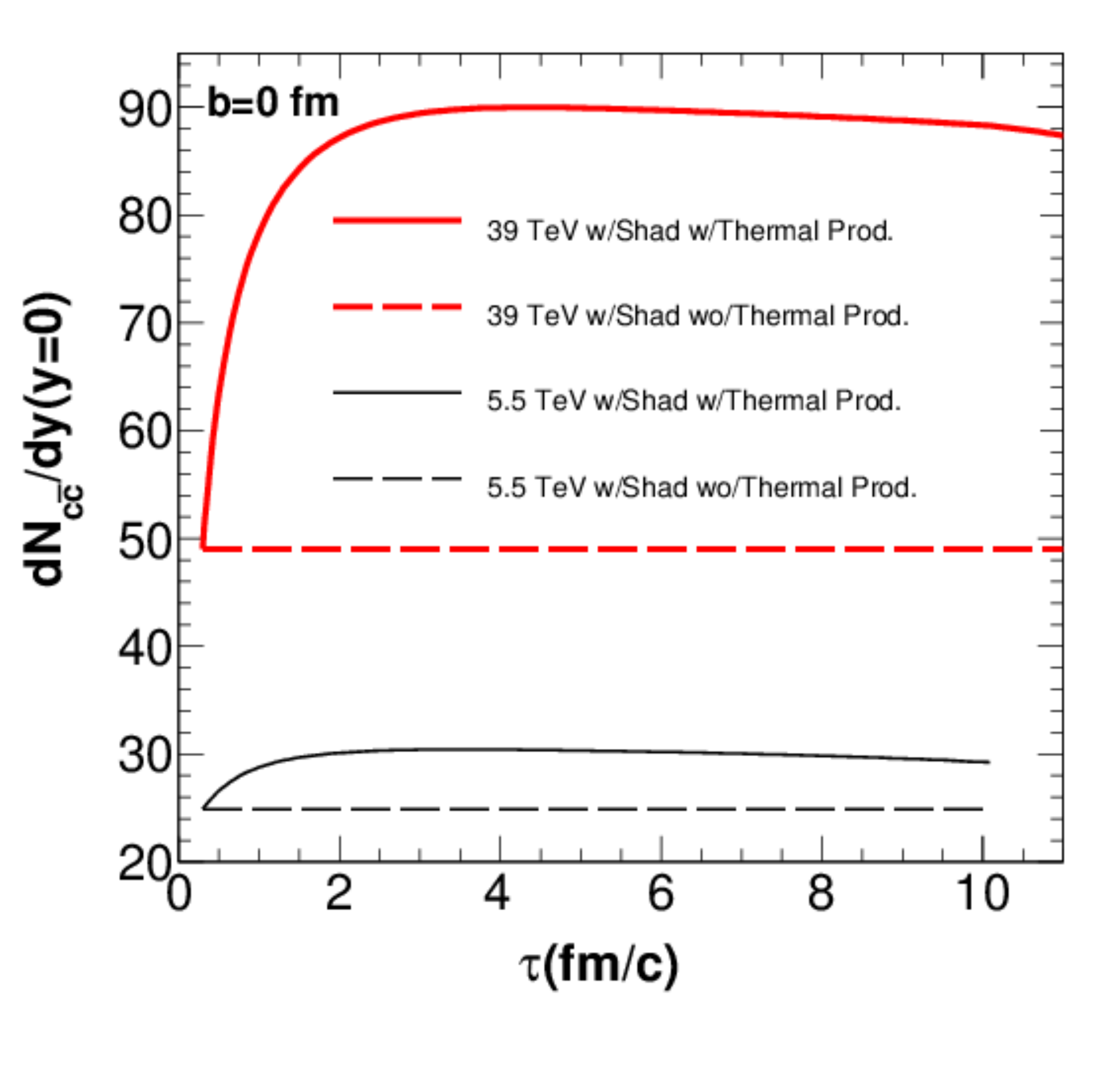}
\hfill
\includegraphics[width=0.54\textwidth]{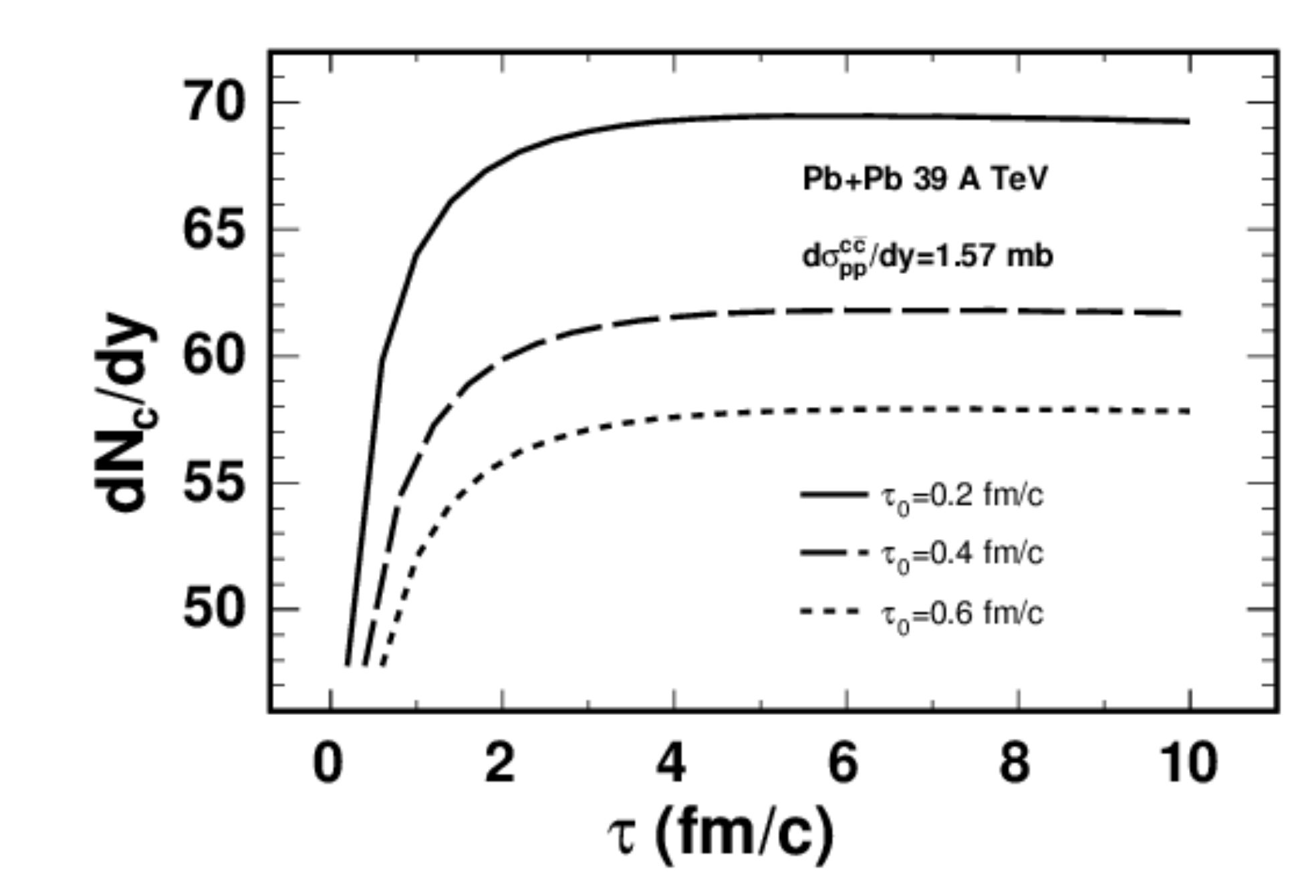}
\caption{Time-evolution of the $c\overline c$ yield
  (per unit of rapidity at midrapidity) for central Pb--Pb collisions
  at $\sqrtsNN=39$~TeV~\cite{Zhou:2016wbo,Liu:2016zle}.}
\label{fig:thermalcharm}
\end{center}
\end{figure}

Figure~\ref{fig:thermalcharm} shows the predictions~\cite{Liu:2016zle,Zhou:2016wbo} for the time-dependence of the $c\overline c$
rapidity density at mid-rapidity in central Pb--Pb collisions at FCC. The value at the initial time
$\tau_0$ corresponds to that of the initial hard-scattering cross section ($\tau\sim 1/2m_{c}$).
Both calculations show a rapid increase
after $\tau_0$ with 
a final yield that is larger by up to 80\% than that corresponding to the initial production.
The increase obtained for top LHC energy 
is of about 15\%.
The thermal charm production would result in an enhancement of charmed hadron 
production at very low $p_{\rm T}$, with respect to the expectation
from binary scaling of the production in pp collisions, after
correction for the nuclear initial-state effects (PDF modification),
that should be measured using proton--nucleus collisions. 
This enhancement provides valuable insights on the QGP temperature evolution.

The abundance of charm quarks also has an effect on the QGP equation
of the state: 
the inclusion of the charm quark in the lattice QCD calculations results in a sizeable 
increase of $P/T^4 \propto n_{\rm d.o.f.}$ for $T>400$~MeV, as
discussed in the CERN Yellow Report~\cite{Dainese:2016gch}.  

\subsubsection{Quarkonium suppression and (re)generation}

\begin{figure}[!t]
\begin{center}
\includegraphics[width=0.42\textwidth]{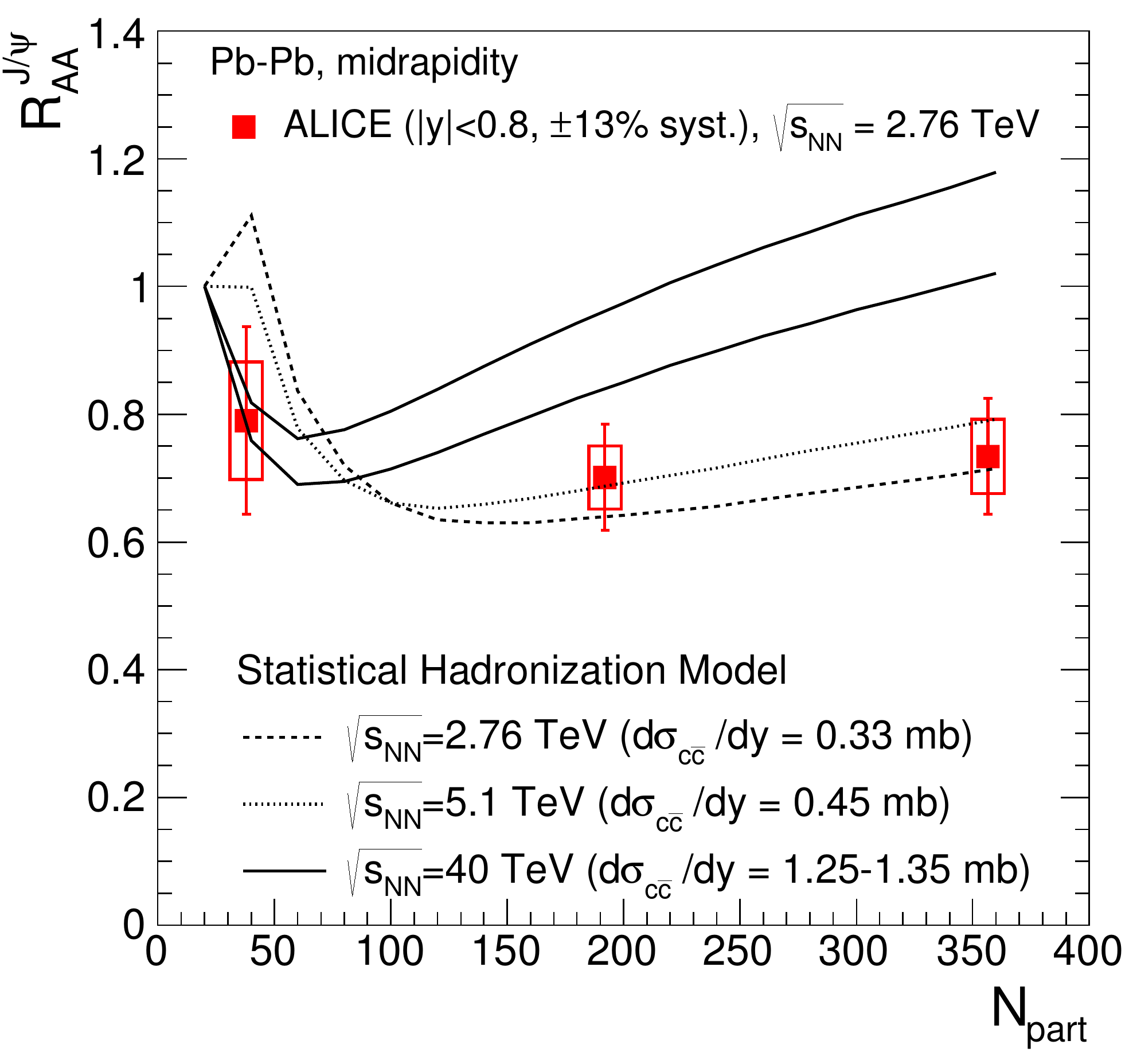}
\hfill
\includegraphics[width=0.42\textwidth]{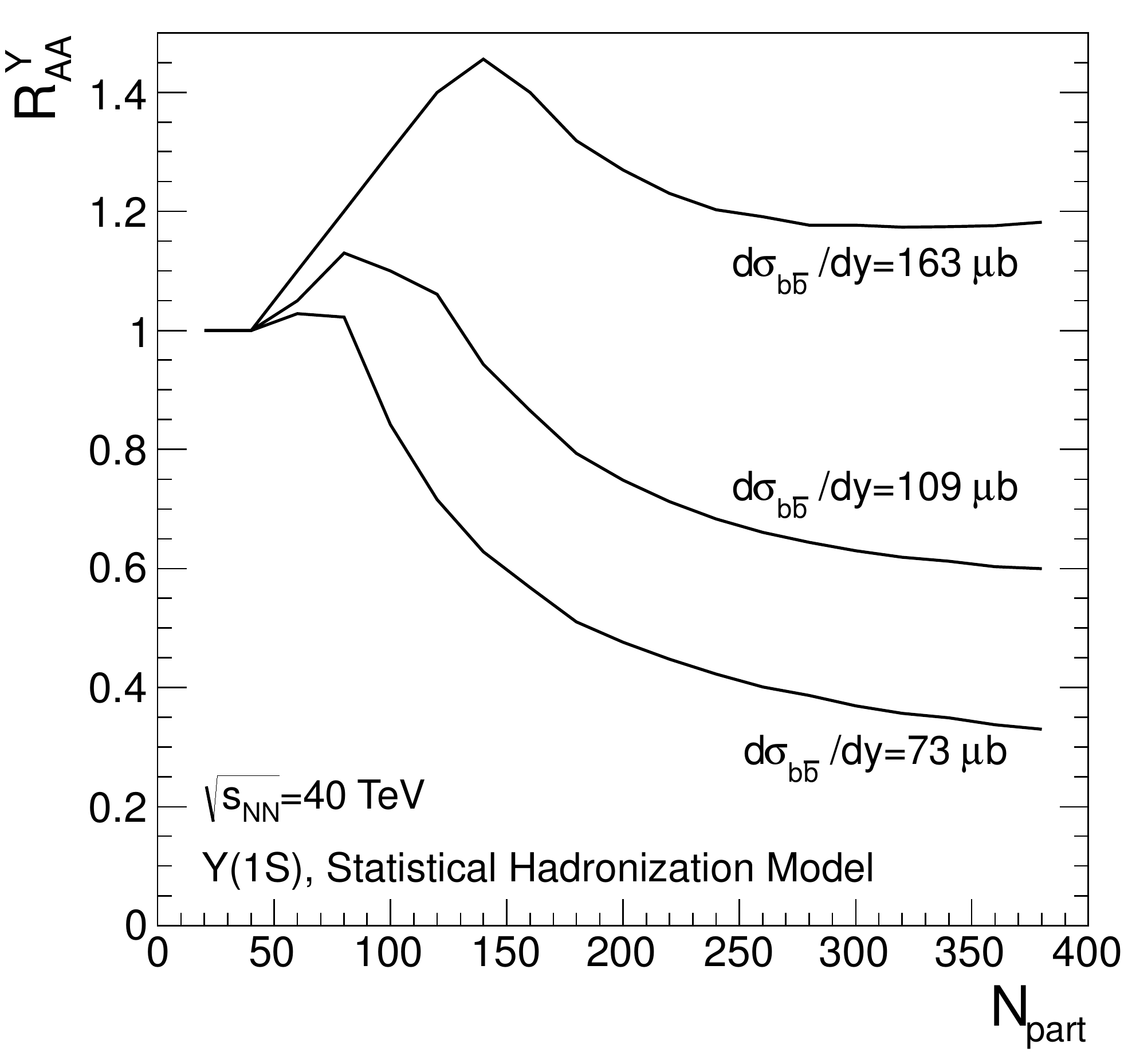}
\caption{Nuclear modification factor $\RAA$ of J$/\psi$ (left) 
and $\Upsilon(1S)$ mesons (right)
at LHC and FCC energies as a function of the collision centrality
(number of participant nucleons)~\cite{Andronic:2011yq}.}
\label{fig:oniaRAA}
\end{center}
\end{figure}

The measurements of the nuclear modification factor of J$/\psi$ at the LHC~\cite{Adam:2015isa,Adam:2015rba,Chatrchyan:2012np} 
are described by models that include dissociation caused by
colour-charge screening and a contribution of recombination
(usually denoted (re)generation) from deconfined $c$ and $\overline c$
quarks in the QGP~\cite{Liu:2009nb,Zhao:2011cv,Andronic:2011yq}. 
The (re)generation contribution is expected to be proportional to
the rapidity density of $c\overline c$ pairs in the QGP. 
Therefore, it is predicted to be much larger
at FCC than LHC energies, as a consequence of the larger hard-scattering
production cross section of $c\overline c$ pairs and the possible 
sizeable thermal production discussed in the previous
section.
This could lead to the observation of an enhancement of J$/\psi$
production with respect to binary scaling of the yield in pp
collisions, i.e.\,$R_{\rm AA}>1$, which would be a striking evidence of 
$c\overline c$ recombination from a deconfined QGP.
Figure~\ref{fig:oniaRAA} (left) shows the predicted J$/\psi$ $\RAA$ at FCC
energy, as obtained with the Statistical Hadronization
Model. 
The model predicts $\RAA(\pt>0)>1$ in central collisions and an
increase of about 40\% with respect to top LHC energy.

The measurement of $\Upsilon$ production would be particularly interesting
at the high
energies and temperatures reached at the FCC.
The LHC data are consistent with a scenario in which the excited
states 2S and 3S are partially or totally suppressed by colour
screening, while the 1S, which is the most tightly bound state, has no
or little direct melting. Its suppression by about 50\% can be
attributed to the lack of feed-down from the (melted) higher states
(see e.g.\,Ref.~\cite{Andronic:2015wma} for a recent review).
At FCC energies, on the one hand, the temperature could be large
enough to determine a full melting even of the tightly-bound 1S state,
on the other hand the large abundance of $b\overline b$ pairs in the
QGP could induce substantial $\Upsilon$ (re)generation.
The possibly large effect of (re)generation of bottomonia from $b$
and $\overline b$ quarks 
is illustrated by the prediction of the Statistical
Hadronization Model~\cite{Andronic:2011yq} for the $\RAA$ of $\Upsilon$(1S) as a function of centrality, shown in the right panel of  
Fig.~\ref{fig:oniaRAA}. 
The predictions are calculated for values of
$d\sigma_{b\overline b}/dy$ in nucleon--nucleon collisions ranging 
from 73 to 163~$\mu$b, as obtained at NLO~\cite{Mangano:1991jk},
which result in 15--40 $b\overline b$ pairs
in central Pb--Pb collisions.
The resulting $\Upsilon$(1S) $\RAA$ in central Pb--Pb collisions is predicted
to range between 0.3 and 1.2.
The role of the two effects ---degree of survival of initial bottomonia and contribution of
(re)generation--- could be separated by means of precise measurements of
the $b\overline b$ cross section
and of the $B$ meson and $\Upsilon$ $\RAA$ and elliptic flow $v_2$ 
(the regenerated $\Upsilon$ states could exhibit a $v_2$ such that $0<v_2^{\Upsilon}<v_2^B$).

\section{Nuclear PDF measurements and search for non-linear QCD (parton saturation)}
\label{sec:HI_smallx}

\newcommand{\pizero}{\ensuremath{\pi^{0}}}
\newcommand{\ptjet}{\ensuremath{p_\mathrm{T,jet}}}

\subsection{Studies in hadronic p--A and A--A collisions}
\label{sec:HI_smallxhadro}

\begin{figure}[t]
\centering
\includegraphics[width=0.35\textwidth]{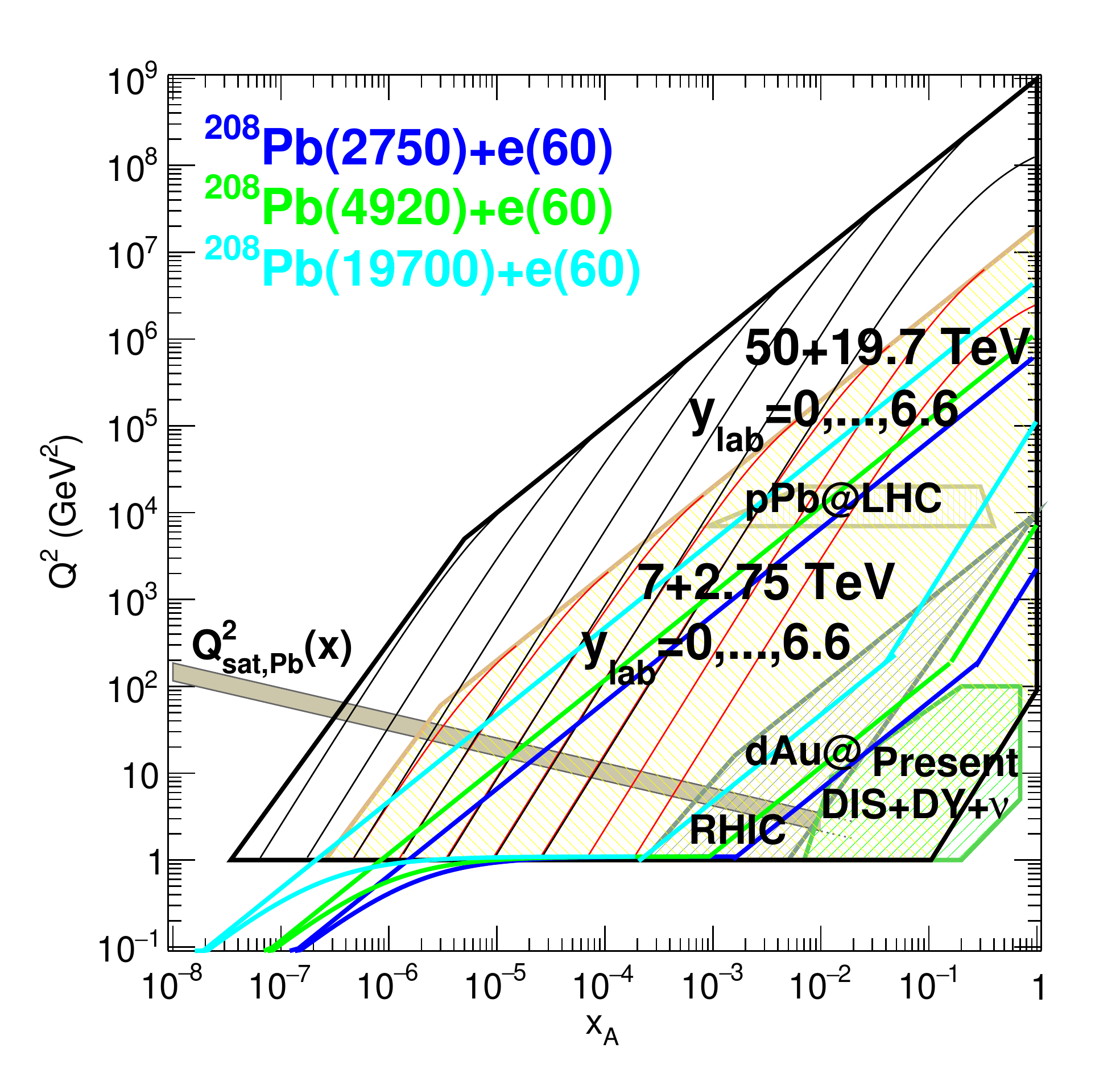}\hskip 2cm\includegraphics[width=0.35\textwidth]{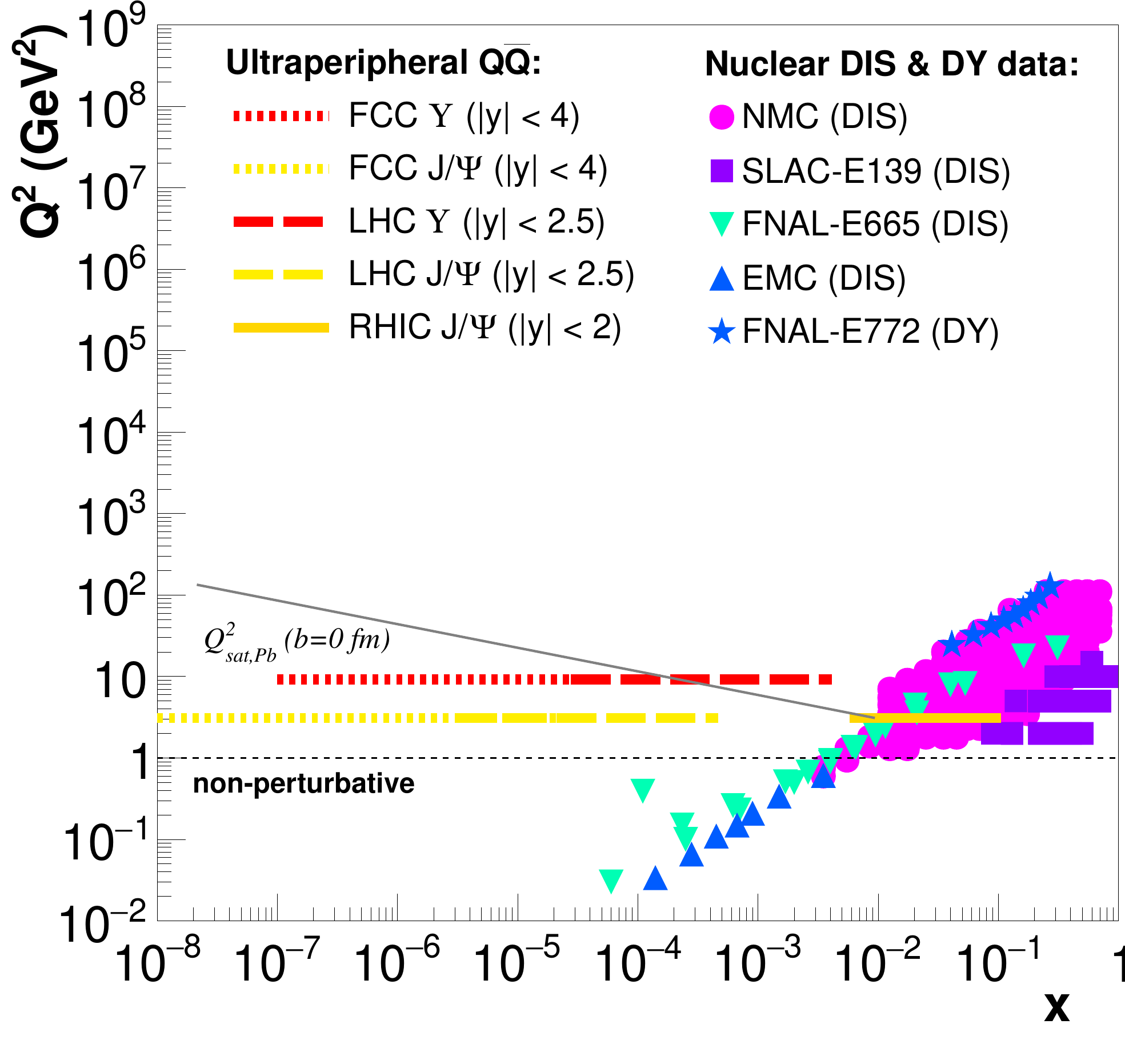}
\caption{Regions of the $x$--$Q^2$ plane covered 
with nuclear DIS and Drell-Yan data and with p--A and e--A collisions at RHIC, LHC, HE-LHC and FCC (left) and exclusive quarkonium photoproduction in $\gamma$--A collisions (right). 
For p--A  at the LHC and the FCC, thin lines correspond to different rapidities in the
laboratory frame $y_{\rm lab} = 0$,1, 2, 3, 4, 5, 6 from right to left, with the left edge defined by $y_{\rm lab} = 6.6$. Values of $Q^2_{\rm S}(x)$ for Pb are shown for illustration.}
\label{fig:kinplane}
\end{figure}

Parton saturation~\cite{Gribov:1984tu,Mueller:1985wy} is based on the idea that standard linear parton branching  leads, at small values of momentum fraction $x$, to parton densities so high that non-linear dynamics (like gluon recombination) becomes important and parton densities are tamed to grow from power-like to logarithmically. Non-linear effects are expected to become important when the density of gluons per unit transverse area exceeds a certain limit, encoded in the saturation momentum $Q_{\rm S}$. 
This regime of QCD is interesting in its own right, to study non-linear dynamics of gluon fields in a controlled environment, but it is also important as an initial state for collisions of nuclei in which a Quark-Gluon Plasma is formed. 
In order to firmly establish the existence of this new high-energy regime of QCD and clarify the validity of the different approaches of perturbative cross sections factorisation and of parton radiation evolution, new kinematic regions must be explored using higher collision energies in order to have a large lever arm in $Q^2$ in a region that, while perturbative, lies inside the saturation domain. The FCC offers such energies and the possibility of combining proton and nuclear beams, as required for a detailed understanding of the mechanism underlying saturation (Fig.~\ref{fig:kinplane}).

There is a strong complementarity between the physics programmes at hadron colliders
and at the proposed electron--hadron colliders (Electron-Ion Collider in the USA~\cite{Accardi:2012qut}, Large Hadron Electron Collider LHeC~\cite{AbelleiraFernandez:2012cc} and FCC-he~\cite{Mangano:2651294}).
With kinematic reach at the TeV scale in the c.m.s.\,(Fig.~\ref{fig:kinplane}, left), these two latter projects are well-positioned to reach conclusive evidence for the existence of a new non-linear regime of QCD. They are clearly complementary with the FCC-hh, providing a precise knowledge on the partonic structure of nucleons and nuclei and on the small-$x$ dynamics.

A full discussion of the observable sensitive to saturation at FCC-hh can found in the CERN Yellow Report~\cite{Dainese:2016gch}. Here we concentrate on two observables, both of which require measurements at forward rapidity $y\approx 4$ or larger: 
single-inclusive spectra of photons and hadrons and azimuthal correlations of di-jets.

\begin{figure}
\centering
\includegraphics[width=0.3\textwidth]{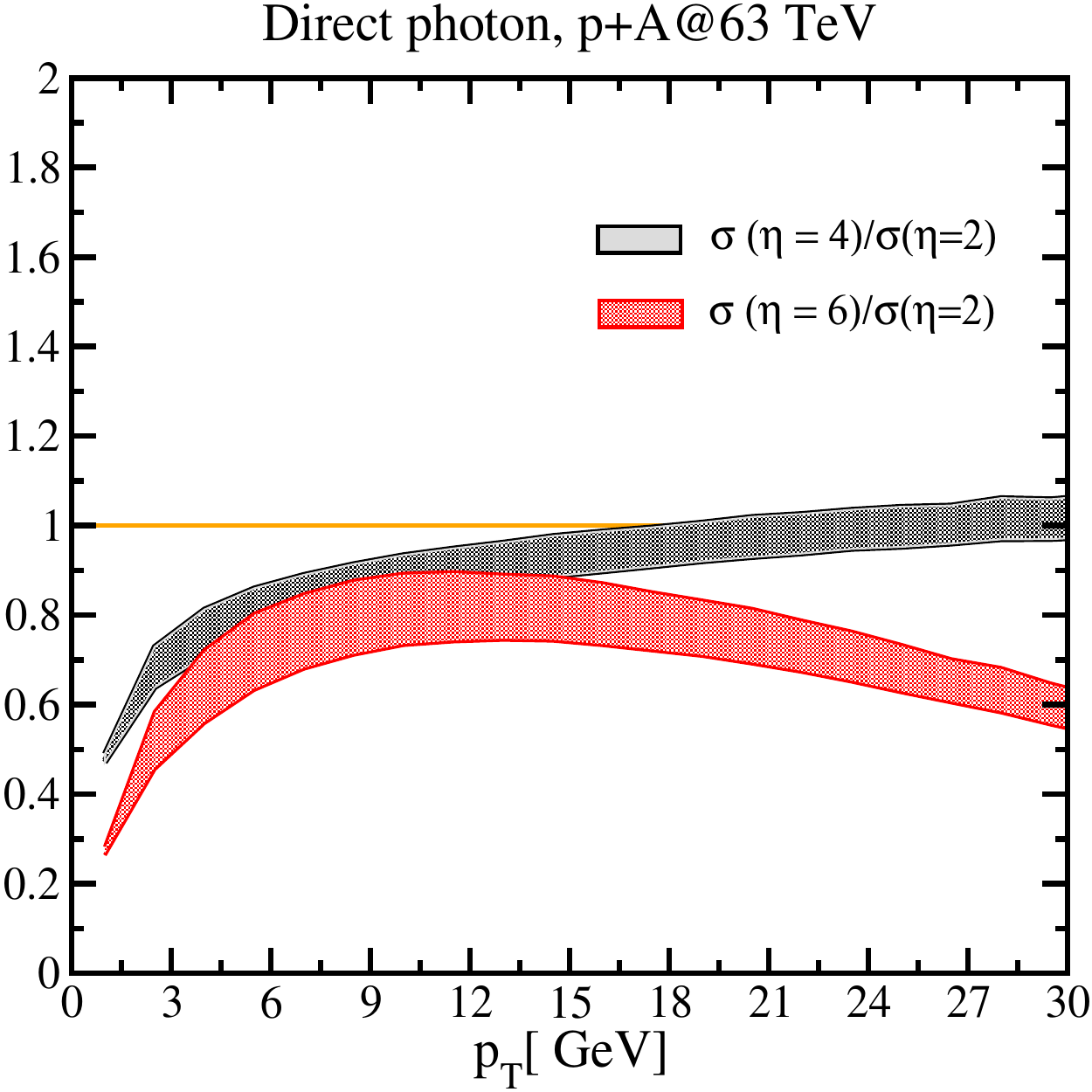}\hfill
\includegraphics[width=0.3\textwidth]{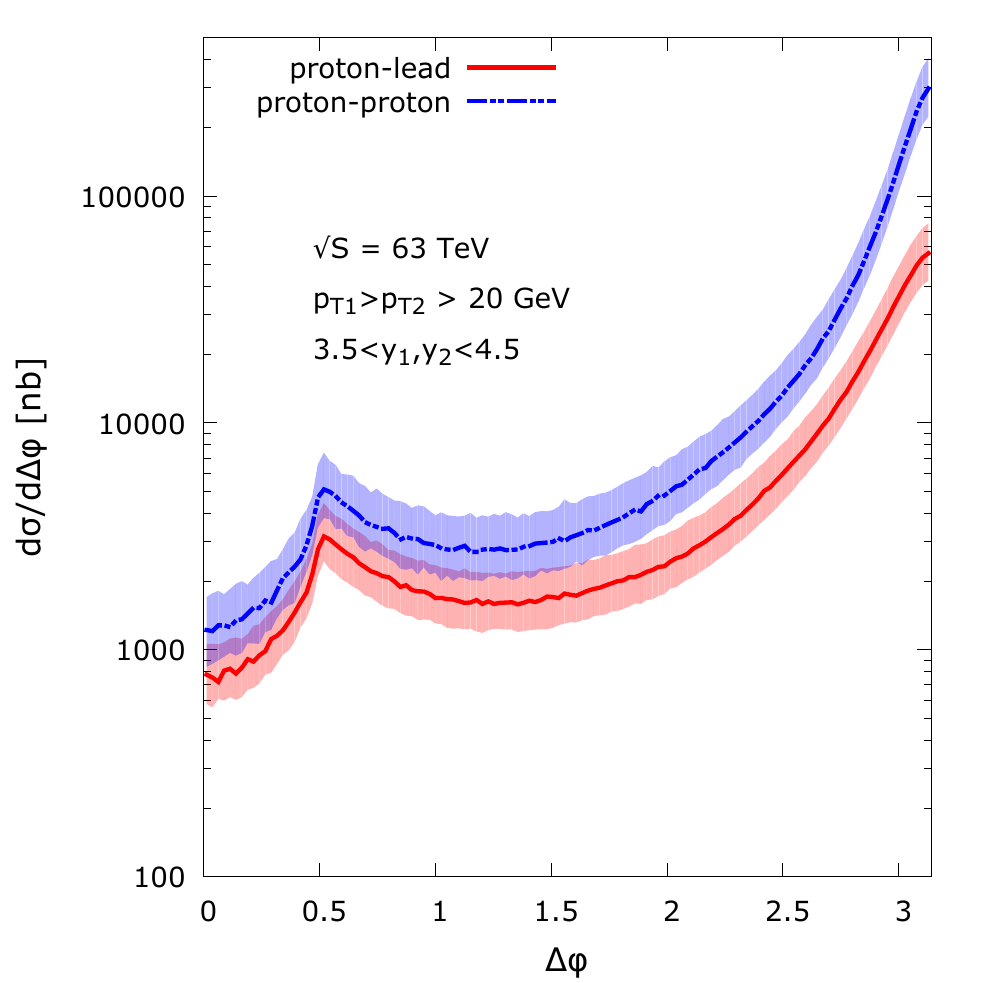}\hfill
\includegraphics[width=0.3\textwidth]{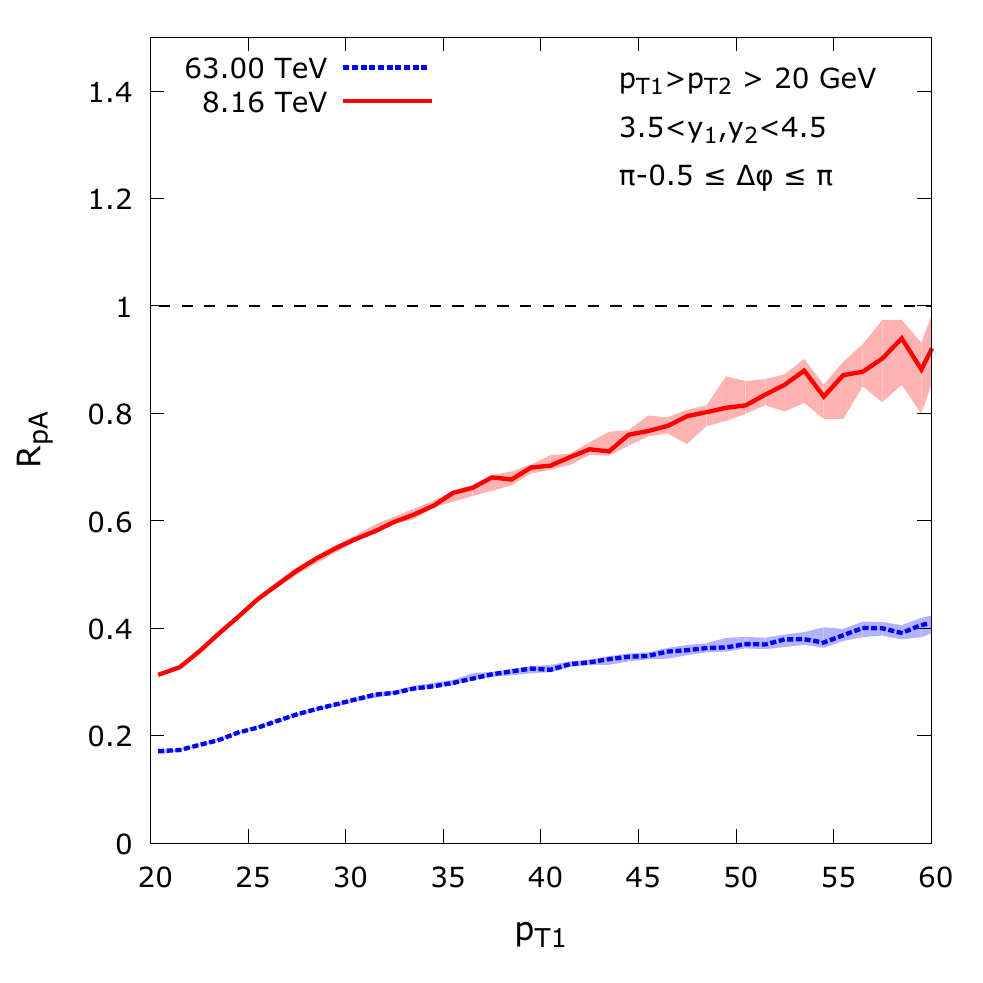}\hfill
\caption{Left: Ratio of direct photons at forward rapidities $\eta=2$, 4, 6 obtained in the CGC formalism in p--Pb collisions~\cite{Rezaeian:2012wa}.
Middle: Forward di-jet $\Delta\varphi$ correlation in p--Pb compared to pp collisions at FCC~\cite{vanHameren:2014lna}. Left: Nuclear modification factor of back-to-back di-jets at LHC and FCC energies.}
\label{fig:marquet}
\end{figure}

 Single-inclusive measurements of charged hadrons, and photons and jets are the simplest observables that probe the gluon density. The two latter  are somewhat safer from a theoretical point of view, as they are free from hadronisation effects to a large extent.
Photons are sensitive to the small-$x$ gluons via the dominant quark--gluon Compton scattering, as well
as hadrons through underlying gluon--gluon scatterings that fragment into hadrons at moderate $\pt$ values.
As shown in Fig.~\ref{fig:kinplane} (left),  measurements at mid-rapidity and $\pt < 10 \GeVc$ at the FCC potentially cover the saturation region with $Q\approx \pt$ and $x$ in the range $10^{-5}$--$10^{-4}$, which is at much lower $x$ and therefore larger gluon density than measurements at the LHC. Forward measurements, e.g.\,at $y \approx 4$, would be even more interesting, as they cover $x \approx 10^{-6}$.
Figure~\ref{fig:marquet} (left) shows the ratios of $\pt$-differential cross sections of direct photons at rapidity 4 and 6 with respect to rapidity 2, calculated in the Color Glass Condensate (CGC) framework~\cite{Rezaeian:2012wa}: saturation is expected to lead to a strong suppression for $\pt<10~\gev/c$ (the high-$\pt$ suppression at $\eta=6$ is an artifact of the calculation).

Forward di-jets offer more potential to experimentally constrain the probed $x$ region, in particular at low \pt. CGC models predict a characteristic suppression of the recoil jet, because (mini-)jets can be produced by scattering a parton off the color field in the nucleus where the recoil momentum is carried by multiple gluons, unlike in a standard (semi-)hard 2-to-2 scattering where all the recoil momentum is carried by a single jet~\cite{Albacete:2010pg,Rezaeian:2012wa}.
First measurements of di-hadron correlations at forward rapidity at RHIC show hints of the predicted suppression of recoil yield. However, the measurements are close to the kinematic limit and other mechanisms have been proposed to explain the suppression.  The FCC has a large potential for recoil measurements of hadrons, photons and jets, probing a broad range in $x$ and $Q^2$, while staying far away from the kinematic limit. 
For di-jets at forward rapidity, 
Fig.~\ref{fig:marquet} shows the expected broadening of the $\Delta\varphi$ distribution in p--Pb with respect to pp collisions (middle), as well as the expected nuclear modification factor for di-jets as a function of the transverse momentum of the leading jet \ptjet{}, in the recoil region (right).
 A clear suppression is visible in the recoil region persisting to much larger $\ptjet > 100$~\GeVc{} than at the LHC, where the suppression is small at $\ptjet \approx 50$~\GeVc. These calculations clearly show sensitivity to parton saturation effects up to high \pt, much larger than $Q_{\rm S}$, as long as the dijet transverse momentum imbalance does not exceed a few times $Q_{\rm S}$.

\begin{figure}[ht]
\centering
\includegraphics[width=0.325\textwidth]{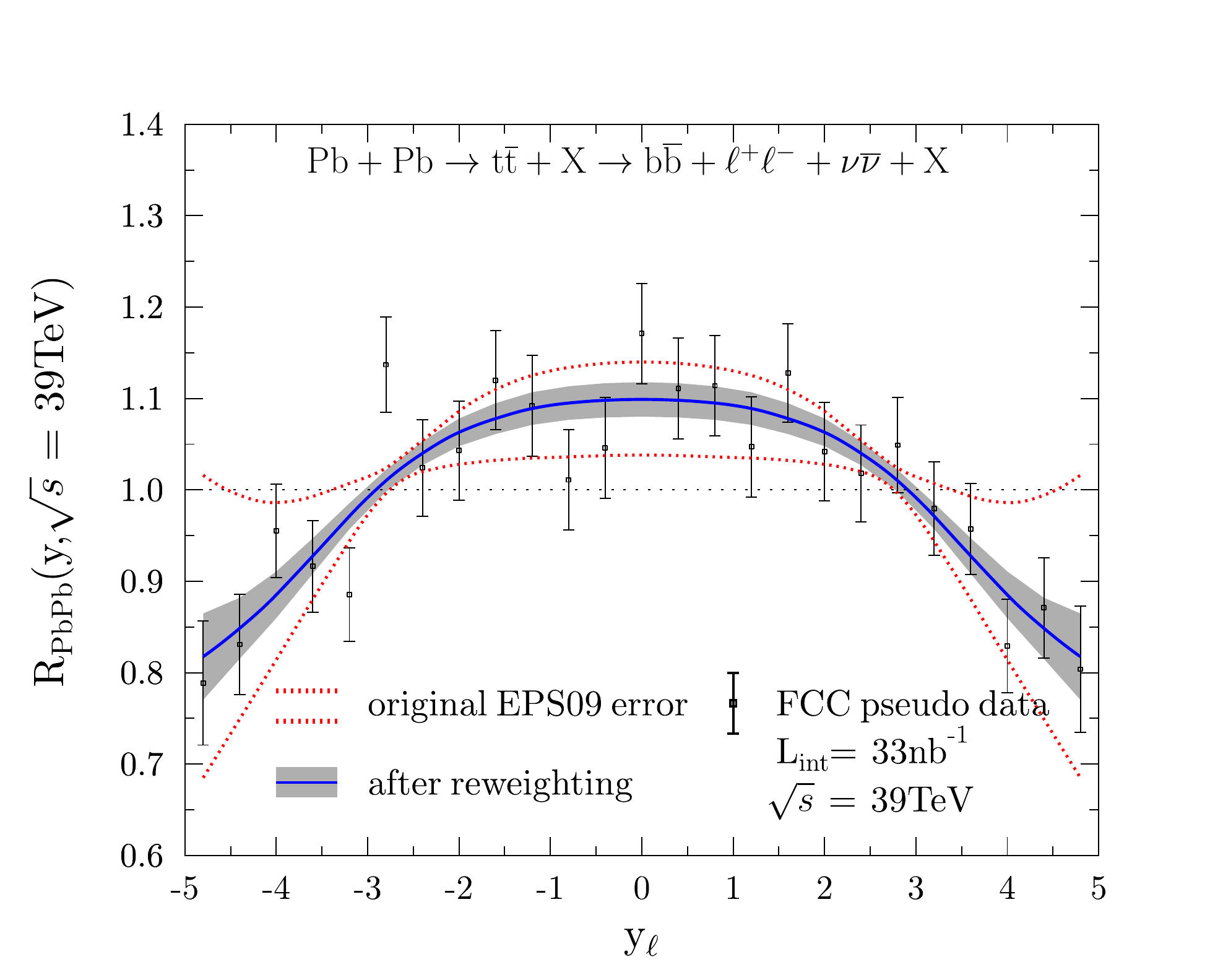}
\includegraphics[width=0.33\textwidth]{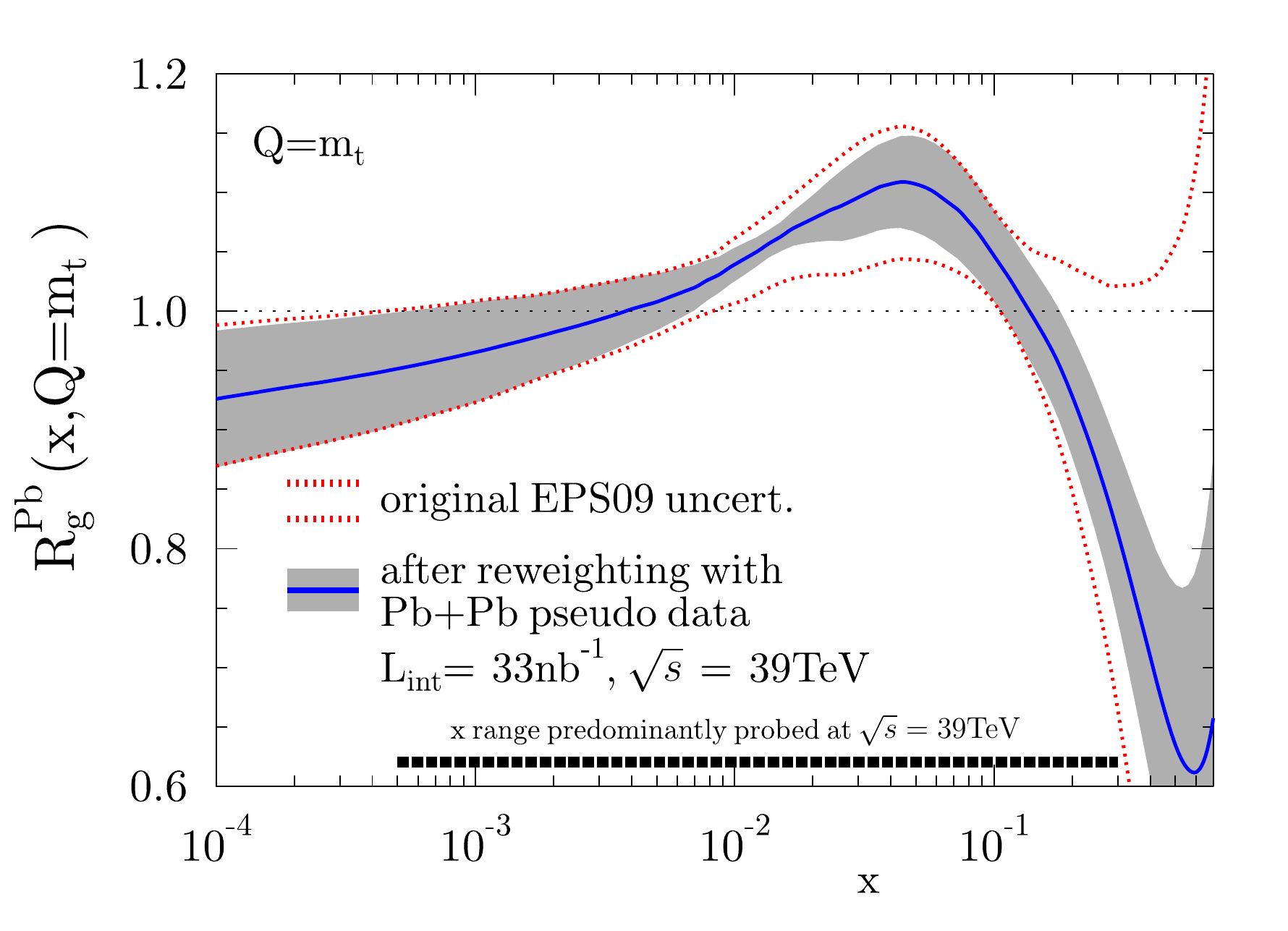}
\includegraphics[width=0.33\textwidth]{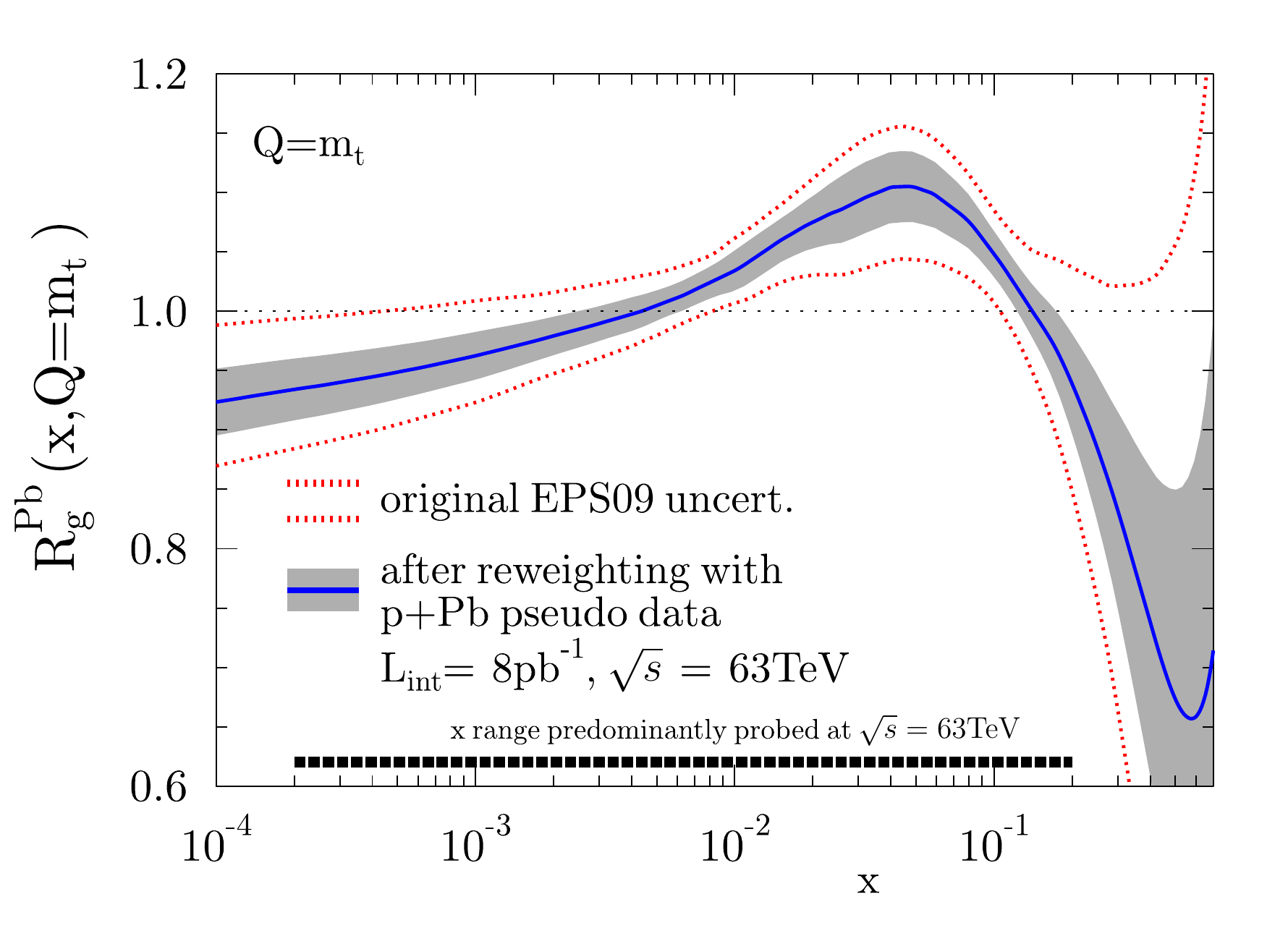}
\caption{Left: FCC pseudodata for the nuclear modification factor for top production in Pb--Pb collisions. Middle and right: Original EPS09 gluon nuclear modification  at $Q=m_{\rm top}$ and 
  estimated improvement obtained by reweighting using the Pb--Pb (middle panel) and p--Pb (right panel). The figures are adapted from Ref.~\cite{dEnterria:2015mgr}.} 
\label{fig:tnPDF}
\end{figure}

Several possibilities exist for constraining nPDFs at large $Q^2$ at the FCC-hh~\cite{Dainese:2016gch}:
$W$ and $Z$ bosons, di-jets, top quarks. Here we will focus on the latter case, because it is a unique 
opportunity offered by the FCC-hh energy increase with respect to the LHC, and it offers 
the possibility to access in experimentally and theoretically well-controlled fashion the 
nuclear PDF at the unprecented large scale $Q=m_{\rm top}$~\cite{dEnterria:2015mgr,dEnterria:2017jyt}. 
To estimate the impact that the FCC would have on nuclear gluon densities, the computed top-pair cross sections in
pp, p--Pb and Pb--Pb with analysis cuts~\cite{Dainese:2016gch} have been binned in the rapidity $y_\ell$ of the $t\to W\to \ell$ decay leptons. 
Figure~\ref{fig:tnPDF} (left) shows Pb--Pb pseudodata for the expected nuclear modification
factors of the decay leptons. 
The effects that these, and the p--Pb, pseudodata would have in the EPS09~\cite{Eskola:2009uj} 
global fit of  nuclear PDFs are quantified via the
Hessian reweighting technique~\cite{Paukkunen:2014zia} and shown in the middle and right panels of the figure. The nPDF uncertainties are reduced by more than
50\% over a large range of $x$.

\subsection{Exclusive photoproduction of heavy quarkonia}
\label{sec:HI_upc}

Charged particles accelerated at high energies generate electromagnetic fields that can be considered as quasireal $\gamma$ beams of very low virtuality $Q^{2} <
1/R^{2}$, where $R$ is the radius of the charge, i.e.\,$Q^2\approx 0.08$~GeV$^2$ for protons
($R\approx 0.7$~fm), and $Q^2<4\cdot 10^{-3}$~GeV$^2$ for nuclei
($R_{\rm A}\approx 1.2\,A^{1/3}$~fm). 
Given that the photon flux
scales with the square of the emitting charge ($Z^2$), the emission of quasireal photons from the Pb-ion is
strongly enhanced compared to that from proton beams (although the
latter feature larger photon energies). 
The maximum centre-of-mass energies  of photon-induced
interactions in ``ultraperipheral'' collisions (UPCs) of proton and lead (Pb)
beams ---occurring at impact parameters larger
than the sum of their radii--- at the FCC are  $W_{\gamma {\rm
    p}}^{\rm max} =10$~TeV and $W_{\gamma A}^{\rm max} =7$~TeV.

In exclusive photoproduction of vector mesons in UPCs of protons or ions
the gluon couples {\em directly}
to the $c$ or $b$ quarks  and the  cross section is proportional to the gluon density {\em squared}, thereby
providing a very clean probe of the gluon density in the ``target''
hadron, with the
large mass of quarkonia providing a hard scale for pQCD calculations.
Their measured cross sections rise steeply with photon-hadron
centre-of-mass energy $W_{\gamma {\rm p}}$, following a power-law
dependence $W_{\gamma {\rm p}}^{\delta}$ with
$\delta=0.7$--$1.2$~\cite{H1:2000,ZEUS:2009}, reflecting the steep rise of the underlying gluon density in the hadrons at 
increasingly lower values of parton fractional momentum $x$.
At the FCC, $\jpsi$ and $\Upsilon$ photoproduction will reach photon-hadron \cm\ energies as large as 
$W_{\gp}\approx 10$~TeV, and thereby probe the gluon density in the proton and nucleus in an
unexplored region down to $x\approx M_{\jpsi,\Upsilon}^2/W_{\gamma {\rm p}}^2\approx
10^{-7}$, at least two orders of magnitude below the range probed at
the LHC (Fig.~\ref{fig:kinplane}, right).

\section{Fixed-target collisions using the FCC proton and lead beams}
\label{sec:HI_fixedtarget}

\newcommand{\glabcms}{\gamma^{\rm lab}_{\rm c.m.s.}}
\newcommand{\blabcms}{\beta^{\rm lab}_{\rm c.m.s.}}
\newcommand{\dylabcms}{\Delta y^{\rm lab}_{\rm c.m.s.}}

The fixed-target mode offers specific advantages compared to the collider mode 
such as accessing the high Feynman $x_F$ domain, achieving high luminosities 
while not impacting on the collider-mode operations, varying the atomic mass of the target almost at will, 
and polarising the target~\cite{Lansberg:2015pra,Dainese:2016gch}.  
The c.m.s.\,energy
in the fixed-target mode is about 200--300~GeV for Pb and proton
beams, respectively. 
The region of central c.m.s.\,rapidities is highly 
boosted at an angle with respect to the beam axis of about one degree 
in the laboratory frame such that the entire backward hemisphere, $y_{\rm c.m.s.}<0$, becomes 
easily accessible with standard experimental techniques at $2<\eta_{\rm lab} <6$. 
A fixed-target setup would allow one to significantly extend a large number of measurements  
carried out at RHIC towards very backward rapidities with much larger luminosities.
As it was discussed in~\cite{Lansberg:2015pra}, RHIC luminosities in $pp$ collisions at 200~GeV are limited and
could not allow for the study of vector boson production close to
threshold which probes
the large $x$ content in the proton and nucleus, 0.7 and above. 
These studies become possible with fixed-target collisions at FCC. 
With a longitudinally polarised target, vector boson production gives access
to (anti)quark helicity distributions in the proton at very large $x$. With deuterium and helium targets, measurements 
can also be carried out on the neutron. Using a transversely polarised target allows one
to access transverse-momentum dependent distributions (TMDs) which are connected to the orbital angular momentum carried
by the partons at larger scales than with any other facilities.

\bibliographystyle{report}
\bibliography{bib/intro,bib/machine,bib/heavyflavour,bib/top,bib/hardxsections,bib/quenching,bib/smallxhadro,bib/upc,bib/soft,bib/fixedtarget,bib/gammagamma}

\providecommand{\href}[2]{#2}\begingroup\raggedright\begin{thebibliography}{10}

\bibitem{Benedikt:2651300}
M.~Benedikt, M.~Capeans~Garrido, F.~Cerutti, B.~Goddard, J.~Gutleber, J.~M.
  Jimenez, M.~Mangano, V.~Mertens, J.~A. Osborne, T.~Otto, J.~Poole,
  W.~Riegler, D.~Schulte, L.~J. Tavian, D.~Tommasini, and F.~Zimmermann, {\em
  {Future Circular Collider - Vol. 3: The Hadron Collider (FCC-hh)}\/},  Tech.
  Rep. CERN-ACC-2018-0058, CERN, Geneva, Dec, 2018.
\newblock \url{https://cds.cern.ch/record/2651300}.
\newblock Submitted for publication to Eur. Phys. J. ST.

\bibitem{Mangano:2651294}
M.~Mangano, P.~Azzi, M.~Benedikt, A.~Blondel, D.~A. Britzger, A.~Dainese,
  M.~Dam, J.~de~Blas, D.~Enterria, O.~Fischer, C.~Grojean, J.~Gutleber,
  C.~Gwenlan, C.~Helsens, P.~Janot, M.~Klein, U.~Klein, M.~P. Mccullough,
  S.~Monteil, J.~Poole, M.~Ramsey-Musolf, C.~Schwanenberger, M.~Selvaggi,
  F.~Zimmermann, and T.~You, {\em {Future Circular Collider- Vol. 1: Physics
  Opportunities}\/},  Tech. Rep. CERN-ACC-2018-0056, CERN, Geneva, Dec, 2018.
\newblock \url{https://cds.cern.ch/record/2651294}.
\newblock Submitted for publication to Eur. Phys. J. C.

\bibitem{Citron:2018lsq}
Z.~Citron et al., {\em {Future physics opportunities for high-density QCD at
  the LHC with heavy-ion and proton beams}\/},  in {\em {HL/HE-LHC Workshop:
  Workshop on the Physics of HL-LHC, and Perspectives at HE-LHC Geneva,
  Switzerland, June 18-20, 2018}}.
\newblock 2018.
\newblock
\href{http://arxiv.org/abs/1812.06772}{{\tt arXiv:1812.06772 [hep-ph]}}.
\newblock

\bibitem{Dainese:2016gch}
A.~Dainese et al., {\em {Heavy ions at the Future Circular Collider}\/},
  \href{http://dx.doi.org/10.23731/CYRM-2017-003.635}{CERN Yellow Report (2017)
  no.~3, 635--692},
\href{http://arxiv.org/abs/1605.01389}{{\tt arXiv:1605.01389 [hep-ph]}}.

\bibitem{Schaumann:2015fsa}
M.~Schaumann, {\em {Potential performance for Pb-Pb, p-Pb and p-p collisions in
  a future circular collider}\/},
  \href{http://dx.doi.org/10.1103/PhysRevSTAB.18.091002}{Phys. Rev. ST Accel.
  Beams {\bf 18} (2015) no.~9, 091002},
\href{http://arxiv.org/abs/1503.09107}{{\tt arXiv:1503.09107
  [physics.acc-ph]}}.

\bibitem{Campbell:2010ff}
J.~M. Campbell and R.~K. Ellis, {\em {MCFM for the Tevatron and the LHC}\/},
  \href{http://dx.doi.org/10.1016/j.nuclphysbps.2010.08.011}{Nucl. Phys. Proc.
  Suppl. {\bf 205-206} (2010)  10--15},
\href{http://arxiv.org/abs/1007.3492}{{\tt arXiv:1007.3492 [hep-ph]}}.

\bibitem{Czakon:2013goa}
M.~Czakon, P.~Fiedler, and A.~Mitov, {\em {Total Top-Quark Pair-Production
  Cross Section at Hadron Colliders Through $O(\alpha^4_S)$}\/},
  \href{http://dx.doi.org/10.1103/PhysRevLett.110.252004}{Phys. Rev. Lett. {\bf
  110} (2013)  252004},
\href{http://arxiv.org/abs/1303.6254}{{\tt arXiv:1303.6254 [hep-ph]}}.

\bibitem{dEnterria:2016ids}
D.~d'Enterria and A.~M. Snigirev, {\em {Triple parton scatterings in
  high-energy proton-proton collisions}\/},
  \href{http://dx.doi.org/10.1103/PhysRevLett.118.122001}{Phys. Rev. Lett. {\bf
  118} (2017)  122001},
\href{http://arxiv.org/abs/1612.05582}{{\tt arXiv:1612.05582 [hep-ph]}}.

\bibitem{dEnterria:2015mgr}
D.~d'Enterria, K.~Krajcz\'ar, and H.~Paukkunen, {\em {Top-quark production in
  proton-nucleus and nucleus-nucleus collisions at LHC energies and beyond}\/},
   \href{http://dx.doi.org/10.1016/j.physletb.2015.04.044}{Phys. Lett. {\bf
  B746} (2015)  64--72},
\href{http://arxiv.org/abs/1501.05879}{{\tt arXiv:1501.05879 [hep-ph]}}.

\bibitem{Baskakov:2015nxa}
A.~V. Baskakov, E.~E. Boos, L.~V. Dudko, I.~P. Lokhtin, and A.~M. Snigirev,
  {\em {Single top quark production in heavy ion collisions at energies
  available at the CERN Large Hadron Collider}\/},
  \href{http://dx.doi.org/10.1103/PhysRevC.92.044901}{Phys. Rev. {\bf C92}
  (2015)  044901},
\href{http://arxiv.org/abs/1502.04875}{{\tt arXiv:1502.04875 [hep-ph]}}.

\bibitem{dEnterria:2018bqi}
D.~d'Enterria and C.~Loizides, {\em {Final-state interactions of the Higgs
  boson in quark-gluon matter}\/},
\href{http://arxiv.org/abs/1809.06832}{{\tt arXiv:1809.06832 [hep-ph]}}.

\bibitem{Berger:2018mtg}
E.~L. Berger, J.~Gao, A.~Jueid, and H.~Zhang, {\em {Higgs properties revealed
  through jet quenching in heavy ion collisions}\/},
\href{http://arxiv.org/abs/1804.06858}{{\tt arXiv:1804.06858 [hep-ph]}}.

\bibitem{Ghiglieri:2019lzz}
J.~Ghiglieri and U.~A. Wiedemann, {\em {Thermal width of the Higgs boson in hot
  QCD matter}\/},
\href{http://arxiv.org/abs/1901.04503}{{\tt arXiv:1901.04503 [hep-ph]}}.

\bibitem{dEnterria:2017jyt}
D.~d'Enterria, {\em {Top-quark and Higgs boson perspectives at heavy-ion
  colliders}\/},
  \href{http://dx.doi.org/10.1016/j.nuclphysbps.2017.05.053}{Nucl. Part. Phys.
  Proc. {\bf 289-290} (2017)  237--240},
\href{http://arxiv.org/abs/1701.08047}{{\tt arXiv:1701.08047 [hep-ex]}}.

\bibitem{Apolinario:2017sob}
L.~Apolinário, J.~G. Milhano, G.~P. Salam, and C.~A. Salgado, {\em {Probing the
  time structure of the quark-gluon plasma with top quarks}\/},
  \href{http://dx.doi.org/10.1103/PhysRevLett.120.232301}{Phys. Rev. Lett. {\bf
  120} (2018) no.~23, 232301},
\href{http://arxiv.org/abs/1711.03105}{{\tt arXiv:1711.03105 [hep-ph]}}.

\bibitem{CasalderreySolana:2012ef}
J.~Casalderrey-Solana, Y.~Mehtar-Tani, C.~A. Salgado, and K.~Tywoniuk, {\em
  {New picture of jet quenching dictated by color coherence}\/},
  \href{http://dx.doi.org/10.1016/j.physletb.2013.07.046}{Phys. Lett. {\bf
  B725} (2013)  357--360},
\href{http://arxiv.org/abs/1210.7765}{{\tt arXiv:1210.7765 [hep-ph]}}.

\bibitem{Andronic:2015wma}
A.~Andronic et al., {\em {Heavy-flavour and quarkonium production in the LHC
  era: from proton-proton to heavy-ion collisions}\/},
  \href{http://dx.doi.org/10.1140/epjc/s10052-015-3819-5}{Eur. Phys. J. {\bf
  C76} (2016)  107},
\href{http://arxiv.org/abs/1506.03981}{{\tt arXiv:1506.03981 [nucl-ex]}}.

\bibitem{Zhou:2016wbo}
K.~Zhou, Z.~Chen, C.~Greiner, and P.~Zhuang, {\em {Thermal Charm and Charmonium
  Production in Quark Gluon Plasma}\/},
  \href{http://dx.doi.org/10.1016/j.physletb.2016.05.051}{Phys. Lett. {\bf
  B758} (2016)  434--439},
\href{http://arxiv.org/abs/1602.01667}{{\tt arXiv:1602.01667 [hep-ph]}}.

\bibitem{Liu:2016zle}
Y.~Liu and C.-M. Ko, {\em {Thermal production of charm quarks in heavy ion
  collisions at Future Circular Collider}\/},
  \href{http://dx.doi.org/10.1088/0954-3899/43/12/125108}{J. Phys. {\bf G43}
  (2016)  125108},
\href{http://arxiv.org/abs/1604.01207}{{\tt arXiv:1604.01207 [nucl-th]}}.

\bibitem{Adam:2015isa}
{ALICE} Collaboration, J.~Adam et al., {\em {Differential studies of inclusive
  $J/\psi$ and $J/\psi$(2S) production at forward rapidity in Pb-Pb collisions
  at $ \sqrt{s_{\mathrm{NN}}}=2.76 $ TeV}\/},
  \href{http://dx.doi.org/10.1007/JHEP05(2016)179}{JHEP {\bf 05} (2016)  179},
\href{http://arxiv.org/abs/1506.08804}{{\tt arXiv:1506.08804 [nucl-ex]}}.

\bibitem{Adam:2015rba}
{ALICE} Collaboration, J.~Adam et al., {\em {Inclusive, prompt and non-prompt
  $J/\psi$ production at mid-rapidity in Pb-Pb collisions at $\sqrt{s_{\rm
  NN}}$ = 2.76 TeV}\/},  \href{http://dx.doi.org/10.1007/JHEP07(2015)051}{JHEP
  {\bf 07} (2015)  051},
\href{http://arxiv.org/abs/1504.07151}{{\tt arXiv:1504.07151 [nucl-ex]}}.

\bibitem{Chatrchyan:2012np}
{CMS} Collaboration, S.~Chatrchyan et al., {\em {Suppression of non-prompt
  $J/\psi$, prompt $J/\psi$, and $\Upsilon$ in PbPb collisions at $\sqrtsNN$ =
  2.76 TeV}\/},  \href{http://dx.doi.org/10.1007/JHEP05(2012)063}{JHEP {\bf 05}
  (2012)  063},
\href{http://arxiv.org/abs/1201.5069}{{\tt arXiv:1201.5069 [nucl-ex]}}.

\bibitem{Liu:2009nb}
Y.-P. Liu, Z.~Qu, N.~Xu, and P.-f. Zhuang, {\em {$J/\psi$ Transverse Momentum
  Distribution in High Energy Nuclear Collisions at {RHIC}}\/},
  \href{http://dx.doi.org/10.1016/j.physletb.2009.06.006}{Phys. Lett. {\bf
  B678} (2009)  72},
\href{http://arxiv.org/abs/0901.2757}{{\tt arXiv:0901.2757 [nucl-th]}}.

\bibitem{Zhao:2011cv}
X.~Zhao and R.~Rapp, {\em {Medium modifications and production of charmonia at
  {LHC}}\/},  \href{http://dx.doi.org/10.1016/j.nuclphysa.2011.05.001}{Nucl.
  Phys. {\bf A859} (2011)  114}, \href{http://arxiv.org/abs/1102.2194}{{\tt
  arXiv:1102.2194 [hep-ph]}}.

\bibitem{Andronic:2011yq}
A.~Andronic, P.~Braun-Munzinger, K.~Redlich, and J.~Stachel, {\em {The thermal
  model on the verge of the ultimate test: particle production in Pb-Pb
  collisions at the LHC}\/},
  \href{http://dx.doi.org/10.1088/0954-3899/38/12/124081}{J. Phys. {\bf G38}
  (2011)  124081},
\href{http://arxiv.org/abs/1106.6321}{{\tt arXiv:1106.6321 [nucl-th]}}.

\bibitem{Mangano:1991jk}
M.~L. Mangano, P.~Nason, and G.~Ridolfi, {\em {Heavy quark correlations in
  hadron collisions at next-to-leading order}\/},
\href{http://dx.doi.org/10.1016/0550-3213(92)90435-E}{Nucl. Phys. {\bf B373}
  (1992)  295--345}.

\bibitem{Gribov:1984tu}
L.~V. Gribov, E.~M. Levin, and M.~G. Ryskin, {\em {Semihard Processes in
  QCD}\/},
\href{http://dx.doi.org/10.1016/0370-1573(83)90022-4}{Phys. Rept. {\bf 100}
  (1983)  1--150}.

\bibitem{Mueller:1985wy}
A.~H. Mueller and J.-w. Qiu, {\em {Gluon Recombination and Shadowing at Small
  Values of x}\/},
\href{http://dx.doi.org/10.1016/0550-3213(86)90164-1}{Nucl. Phys. {\bf B268}
  (1986)  427}.

\bibitem{Accardi:2012qut}
A.~Accardi et al., {\em {Electron Ion Collider: The Next QCD Frontier}\/},
  \href{http://dx.doi.org/10.1140/epja/i2016-16268-9}{Eur. Phys. J. {\bf A52}
  (2016)  268},
\href{http://arxiv.org/abs/1212.1701}{{\tt arXiv:1212.1701 [nucl-ex]}}.

\bibitem{AbelleiraFernandez:2012cc}
{LHeC Study Group} Collaboration, J.~L. Abelleira~Fernandez et al., {\em {A
  Large Hadron Electron Collider at CERN: Report on the Physics and Design
  Concepts for Machine and Detector}\/},
  \href{http://dx.doi.org/10.1088/0954-3899/39/7/075001}{J. Phys. {\bf G39}
  (2012)  075001},
\href{http://arxiv.org/abs/1206.2913}{{\tt arXiv:1206.2913 [physics.acc-ph]}}.

\bibitem{Rezaeian:2012wa}
A.~H. Rezaeian, {\em {Semi-inclusive photon-hadron production in pp and pA
  collisions at RHIC and LHC}\/},
  \href{http://dx.doi.org/10.1103/PhysRevD.86.094016}{Phys. Rev. {\bf D86}
  (2012)  094016},
\href{http://arxiv.org/abs/1209.0478}{{\tt arXiv:1209.0478 [hep-ph]}}.

\bibitem{Albacete:2010pg}
J.~L. Albacete and C.~Marquet, {\em {Azimuthal correlations of forward
  di-hadrons in d+Au collisions at RHIC in the Color Glass Condensate}\/},
  \href{http://dx.doi.org/10.1103/PhysRevLett.105.162301}{Phys. Rev. Lett. {\bf
  105} (2010)  162301},
\href{http://arxiv.org/abs/1005.4065}{{\tt arXiv:1005.4065 [hep-ph]}}.

\bibitem{vanHameren:2014lna}
A.~van Hameren, P.~Kotko, K.~Kutak, C.~Marquet, and S.~Sapeta, {\em {Saturation
  effects in forward-forward dijet production in p$+$Pb collisions}\/},
  \href{http://dx.doi.org/10.1103/PhysRevD.89.094014}{Phys. Rev. {\bf D89}
  (2014)  094014},
\href{http://arxiv.org/abs/1402.5065}{{\tt arXiv:1402.5065 [hep-ph]}}.

\bibitem{Eskola:2009uj}
K.~J. Eskola, H.~Paukkunen, and C.~A. Salgado, {\em {EPS09: A New Generation of
  NLO and LO Nuclear Parton Distribution Functions}\/},
  \href{http://dx.doi.org/10.1088/1126-6708/2009/04/065}{JHEP {\bf 04} (2009)
  065},
\href{http://arxiv.org/abs/0902.4154}{{\tt arXiv:0902.4154 [hep-ph]}}.

\bibitem{Paukkunen:2014zia}
H.~Paukkunen and P.~Zurita, {\em {PDF reweighting in the Hessian matrix
  approach}\/},  \href{http://dx.doi.org/10.1007/JHEP12(2014)100}{JHEP {\bf 12}
  (2014)  100},
\href{http://arxiv.org/abs/1402.6623}{{\tt arXiv:1402.6623 [hep-ph]}}.

\bibitem{H1:2000}
{H1} Collaboration, A.~Atkas et al., {\em Elastic photoproduction of $J/\psi$
  and $\Upsilon$ mesons at HERA\/},
  \href{http://dx.doi.org/10.1016/S0370-2693(00)00530-X}{Phys. Lett. B {\bf
  483} (2000)  23}, \href{http://arxiv.org/abs/hep-ex/0003020}{{\tt
  arXiv:hep-ex/0003020 [hep-ex]}}.

\bibitem{ZEUS:2009}
{ZEUS} Collaboration, S.~Chekanov et al., {\em Exclusive photoproduction of
  $\Upsilon$ mesons at HERA\/},
  \href{http://dx.doi.org/10.1016/physlettb.2009.07.066}{Phys. Lett. B {\bf
  680} (2009)  4}, \href{http://arxiv.org/abs/0903.4205}{{\tt arXiv:0903.4205
  [hep-ex]}}.

\bibitem{Lansberg:2015pra}
J.~P. Lansberg, R.~E. Mikkelsen, and U.~I. Uggerhoej, {\em {Near-threshold
  production of $W^\pm$, $Z^0$ and $H^0$ at a fixed-target experiment at the
  future ultra-high-energy proton colliders}\/},
  \href{http://dx.doi.org/10.1155/2015/249167}{Adv. High Energy Phys. {\bf
  2015} (2015)  249167},
\href{http://arxiv.org/abs/1507.01438}{{\tt arXiv:1507.01438 [hep-ex]}}.

\end{thebibliography}\endgroup


\end{document}